\documentclass[preprint]{JHEP} 
\usepackage{epsfig}
\usepackage{amsmath}
\usepackage{amssymb,amsfonts}

\newcommand{\Comp}{\mathbb{C}}
\newcommand{\Eb}{\mathbb{E}}
\newcommand{\Hb}{\mathbb{H}}
\newcommand{\Tb}{\mathbb{T}}

\def\ol{\overline}

\newcommand{\irrep}[1]{\ensuremath{\boldsymbol{#1}}}
\newcommand{\eis}[3]{~\ensuremath{{\cal E}^{#1}_{\irrep{#2};#3}}}

\newcommand{\Tr}{{\rm Tr}}
\newcommand{\Pf}{{\rm Pf}}
\newcommand{\Zint}{\mathbb{Z}}

\newcommand{\Real}{\mathbb{R}}
\newcommand{\Q}{\mathbb{Q}}

\newcommand{\E}{{\cal E}}

\newcommand{\F}{{\cal F}}

\renewcommand{\a}{\alpha}
\renewcommand{\b}{\beta}
\newcommand{\g}{\gamma}
\newcommand{\w}{\omega}

\newcommand{\pa}{\partial}
\newcommand{\nn}{\nonumber}

\def\wt{\widetilde}
\def\d{\partial}
\def\rank{\mbox{rank}}

\def\sss{\scriptscriptstyle}

\title{Minimal representations, spherical vectors, \\
and exceptional theta series} 

\author{David Kazhdan\\
Dept of Mathematics, Harvard University, Cambridge, MA 02138, USA\\
Email: \email{kazhdan@math.harvard.edu}}

\author{Boris Pioline\footnote{On leave of absence 
from Jefferson Physical Laboratory, 
Harvard University, Cambridge, MA 02138, USA}\\
LPTHE, Universit{\'e}s Paris VI \& VII, \\
Bo\^{\i}te 126, Tour 16, 4 place Jussieu, 75252
Paris, FRANCE
\\
Email: \email{pioline@lpthe.jussieu.fr}}

\author{Andrew Waldron\footnote{On leave of absence from Dept. of
Mathematics, UC Davis, CA 95616.}\\
Physics Department, Brandeis University, Waltham, MA 02454, USA\\
Email: \email{wally@brandeis.edu}}

\preprint{\hepth{0107222}\\
BRX-TH493\\HUTP-01/A036\\LPTHE/01-49}  

\abstract{Theta series for exceptional groups have
been suggested as a possible description of the eleven-dimensional 
quantum supermembrane. We present explicit formulae
for these automorphic forms whenever the underlying Lie group $G$
is split (or complex) and simply laced.
Specifically, we review and construct explicitly 
the minimal representation of $G$, generalizing the Schr\"odinger
representation of symplectic groups.
We compute the spherical vector in this representation,
i.e. the wave function invariant under the maximal compact subgroup,
which plays the role of the summand in the automorphic theta series.
We also determine the spherical vector over the complex field.
We outline how the spherical vector over the $p$-adic number fields
provides the summation measure in the theta series, postponing
its determination to a sequel of this work.
The simplicity of our result is suggestive of a new Born-Infeld-like 
description of the membrane where U-duality is realized non-linearly. 
Our results may also be used in constructing quantum mechanical systems 
with spectrum generating symmetries.}

\dedicated{To appear in Commun. Math. Phys.}

\enlargethispage{1cm}
\begin{document}

\section{Introduction}
Despite considerable insights afforded by dualities,
the fundamental degrees of freedom of M-theory remain elusive. 
Recently the r\^ole of the eleven-dimensional supermembrane has been
tested~\cite{Pioline:2001jn} in an attempt to rederive 
toroidally compactified, M-theoretic, supersymmetric four-graviton
scattering amplitudes at 
order $R^4$. 
These amplitudes are known independently on the basis of
supersymmetry and duality, to be given by an Eisenstein series
of the U-duality 
group~\cite{Green:1997tv,Kiritsis:1997em,Pioline:1998mn,Obers:1999um}
(see~\cite{Obers:1998fb} for a review),
but still lack a finite microscopic derivation
(see however \cite{Green} for a discussion of perturbative computations
in eleven-dimensional supergravity).
In analogy with the string one-loop computation, a one-loop 
membrane amplitude 
was constructed as the integral of a modular invariant
partition function on the fundamental domain of a membrane modular group
$Gl(3,\Zint)$. The action of a membrane instanton configuration
with given winding numbers is given by the Polyakov action and
as a working hypothesis the summation measure was taken to be unity. 
A comparison to the exact result showed
that the mass spectrum and the instanton saddle points were correctly
reproduced by this ansatz, but the spectrum multiplicities 
and instanton summation measure were incorrect\footnote{See 
\cite{Sugino:2001iq} for a very recent discussion of the membrane summation
measure.}. The
proposed partition function was therefore not U-duality invariant.
However, a general method to construct invariant partition functions 
was outlined: exceptional theta series should
provide the correct partition function for the BPS membrane on torii.

While theta series for symplectic groups are very common both in
mathematics, e.g., in the study of Riemann surfaces, and physics
where they arise as partition functions of free theories, their
generalization to other groups is not as well understood. One difficulty is
that group invariance requires a generalization
of the standard Poisson resummation formula (i.e., Gaussian integration)
to cubic characters (i.e., ``Airy'' integration). This scenario is clearly
well adapted to the membrane situation, where the Wess-Zumino interaction
is cubic in the brane winding numbers. Since theta series reside 
at the heart of many problems in the theory of automorphic forms,
it would be very desirable from both physical 
and mathematical viewpoints, to have explicit expressions for them.

As outlined in~\cite{Pioline:2001jn}, the construction of theta series for a 
simple non-compact group $G$
requires three main ingredients: (i)~An irreducible
representation of the group in an appropriate space of functions. 
In the symplectic case, this is simply
the Weil representation of the Heisenberg algebra $[p_i,x^j]=-i\delta_i{}^j$,
which gives rise to the Schr\"odinger representation of $Sp(n,\Real)$.
(ii)~A special function $f$, known as the spherical vector,
which is invariant under the maximal compact subgroup $K$ of $G$.
This generalizes the Gaussian character $e^{2\pi i (x^i)^2}$
appearing in the symplectic theta series. (iii)~A distribution 
$\delta$ invariant under an arithmetic subgroup $G(\Zint)\subset G$  
generalizing
the sum with unit weight over integers $x^i\in \Zint$ of the symplectic case.

As for step (i), one observes that for any simple Lie algebra $\mathcal
G$  there exists a unique non-zero minimal conjugacy class  $\mathcal O
\subset \mathcal G$. This nilpotent orbit carries the standard Kirillov-Kostant
symplectic form, whose quantization furnishes a representation of $G$
on the Hilbert space of wave functions on a Lagrangian submanifold 
of~${\mathcal O}$. Its quantization relies heavily on the existence,
discovered by Joseph~\cite{Joseph}, of a unique completely prime 
two-sided ideal J of the enveloping algebra $U(\mathcal G)$ whose
characteristic variety coincides with $\mathcal O \cup \{0\}$.
The obtained representation is minimal, in the sense that
its Gelfand-Kirillov dimension is smallest among all representations,
being equal to half the dimension of $\mathcal O$. 
The minimal representation exists not only for the split real 
group $G(\Real)$, but also for the group $G(F)$
for arbitrary local field F as long as $\mathcal G$ is any simply-laced
split Lie algebra. In the case when  $\mathcal G$ is of the type $D_n$,
the minimal representation can be realized using Howe's theory of dual
pairs ~\cite{Howe}. The general construction was described in
~\cite{Kostant} and~\cite{KazhdanS}, the latter of which we will
closely follow in this work. 
Step (ii) is the main subject of the present paper; we will obtain the
spherical vector for all groups $G(\Real)$ of $A,D,E$ type in the split real
form, using techniques from Eisenstein series ($A_n$), dual pairs ($D_n$) 
and PDE's ($E_{6,7,8}$). A simple generalization will also provide
the spherical vector for the complex group $G(\Comp)$.
As we will see, step~(iii) 
amounts to solving step~(ii) over all $p$-adic number fields ${\mathbb{Q}}_p$
instead of the
reals. Our methods will allow us to obtain the $p$-adic spherical vector
for $A$ and $D$ groups. The exceptional case requires more powerful
techniques, and will be treated in a sequel to this paper~\cite{sasha}.

While this paper is mostly concerned with the mathematical construction
of exceptional theta series, a few words about the physical implications of our
results are in order. First and foremost, we find that a membrane partition
function invariant under both the modular group $Gl(3,\Zint)$ and
the U-duality group $E_{d}(\Zint)$ {\it cannot} be constructed by
summing over the $3d$ membrane winding numbers alone (which confirms
the findings of~\cite{Pioline:2001jn}). Indeed, the dimension
of the minimal representation of the smallest simple group $G$ containing
$Sl(3,\Real)\times E_{d}$ is always bigger than~$3d$. Second,  
we find that the minimal representation of $G$ has a structure quite 
reminiscent of the membrane, but (in the simplest $d=3$ case)
necessitates two new quantum numbers, which would be very interesting
to understand from the point of view of the quantum membrane. In fact,
the form of the spherical vector in this representation,
displayed in equations~\eqref{e563} and~\eqref{e592} below, is very suggestive
of a Born-Infeld-like formulation of the membrane, which would then
exhibit a hidden dynamical $E_{d+2}(\Zint)$ symmetry. 
A more complete physical analysis of these results in the context of 
the eleven-dimensional supermembrane will appear elsewhere. In addition, 
our minimal representation provides the
quantized phase space for quantum mechanical systems with dynamical
non-compact symmetries, which may find a use in M-theory or other contexts.
By choosing one of the compact generators as
the Hamiltonian, one may construct integrable quantum mechanical 
systems with a spectrum-generating exceptional symmetry, 
and the spherical vector we constructed would then give the ground state
wave function.

The organization of this paper is as follows: In Section~2, we 
use the $Sl(2)$ case as a simple example to introduce the main technology.
In Section~3, we review the construction of the
minimal representation for simply-laced groups. Section~4 contains
the new results of this paper; real and complex spherical vectors for
all $A,D,E$ groups (the main formulae may be found in 
equations~\eqref{e49},~\eqref{fdn},~\eqref{efE6},~\eqref{eFE7},~\eqref{efE8},~\eqref{complex} and~\eqref{mould}). 
We close in Section~5 with a preliminary
discussion of the physics interpretation of our formulae. 
Miscellaneous group theoretical data is gathered in the Appendix.

\section{$Sl(2)$ revisited}
As an introduction to our techniques,
let us consider two familiar examples of automorphic forms 
for $Sl(2,\Zint)$. 

\subsection{Symplectic theta series}
Our first example is the standard Jacobi theta series
\begin{equation}
\label{e1}
\theta(\tau)=\tau_2^{1/4} \sum_{m\in\Zint} e^{i\pi  \tau m^2}
= \sum_{m\in\Zint}f_\tau(m)\ ,\qquad 
f_\tau(x)=\tau_2^{1/4}e^{i\pi\tau x^2}\ ,
\end{equation}
where we inserted a power of $\tau_2$ to cancel the modular weight.
As is well known, this series is an holomorphic modular form 
of $Sl(2,\Zint)$ up to a system of phases. 
The invariance under the generator 
$T:\tau\to\tau+2$ is manifest, while the transformation
under $S:\tau\to-1/\tau$ yielding 
\begin{equation}
\label{e2}
\theta(-1/\tau)= \sqrt{i} \;\; \theta(\tau)\, ,
\end{equation}
follows from the Poisson resummation formula,
\begin{equation}
\label{e3}
\sum_{m\in\Zint} f(m)=
\sum_{p\in\Zint} \wt f(p)\, ,\qquad
\wt f(p)\equiv\int dx\, f(x) \ e^{2\pi i p x}\, ,
\end{equation}
applied to the Gaussian kernel $f_\tau(x)$. 
A better understanding of the mechanism behind
the invariance of the theta series~\eqref{e1}
can be gained (see e.g.,~\cite{tata}) by
rewriting it as
\begin{equation}
\label{e4}
\theta(\tau)=\langle \delta, \rho(g_\tau)\cdot f \rangle\ .
\end{equation}
In this symbolic form, $\rho$ is a representation of the double cover 
$\tilde G$ of $Sl(2,\Real)$ 
in the space ${\cal S}$ of Schwartz functions of one variable;
$g_\tau=\begin{pmatrix} 1&\tau_1\\0 & \tau_2\end{pmatrix}/\sqrt{\tau_2}$ 
is an element of $G=Sl(2,\Real)$ 
parameterizing the coset $U(1)\backslash Sl(2,\Real)$ in the Iwasawa gauge;
 $f(x)=e^{-x^2/2}$ is the spherical vector of the representation
$\rho$, {\it i.e.} an element of ${\cal S}$  which is an eigen-vector of the
preimage $\tilde U\subset \tilde G$  of 
the maximal compact subgroup $K=U(1)$ of $G$ corresponding to the basic
character of $\tilde U$; finally,  
$\delta_{\Zint}(x)=\sum_{m\in\Zint}\delta(x-m)$ 
is a distribution in the dual space of ${\cal S}$, invariant under
the action of $Sl(2,\Zint)$. [The inner product 
$\langle\cdot,\cdot\rangle$ is just integration $\int dx$.]
The invariance of $\theta(\tau)$ then follows trivially
from the covariance of the various pieces in~\eqref{e4}.

More explicitly, $\rho$ is the so-called metaplectic representation
\begin{eqnarray}
\label{e5}
&\rho\!&\begin{pmatrix} 1 & t\\ 0 & 1 \end{pmatrix}:
\phi(x) \to e^{i\pi t x^2} \phi(x) \, ,\\
\label{e5.5}
&\rho\!&\begin{pmatrix} e^{-t} & 0\\ 0 & e^{t} \end{pmatrix}:
\phi(x) \to e^{t/2}\phi(e^{t} x ) \, ,\\
\label{e5.61}
&\rho\!&\begin{pmatrix} 0 & -1\\ 1 & 0 \end{pmatrix}:
\phi(x) \to e^{i\pi/4}\wt \phi(-x)\, .
\end{eqnarray}
acting on a function $\phi\in {\cal S}$.
It is easily checked that the defining relation $(ST)^3=1$
holds modulo a phase, and that the generators $S$ and $T$
leave the distribution $\delta$ invariant.
Linearizing~\eqref{e5} and~\eqref{e5.5} yields generators
for the positive root and Cartan elements
\begin{equation}
\label{e6}
E_+= i\pi\,x^2\, , 
\qquad H=\frac{1}{2}\,(x\partial_x+\partial_x x)\, ,
\end{equation}
while the negative root follows by a Weyl reflection
\begin{equation}
\label{e6.5}
E_-=-\rho(S)\cdot E_+ \cdot \rho(S^{-1})=\frac{i}{4\pi}\,\d_x^2\, ,
\end{equation}
and we have the $Sl(2,\Real)$ algebra,
\begin{equation}
\label{e7}
[H,E_\pm]=\pm\,2\, E_\pm\ ,\quad H=[E_+,E_-]\ .
\end{equation}
In this representation, there does not exist a spherical vector strictly 
speaking, since the compact generator $E_+ - E_-$ (recognized as the
Hamiltonian of the harmonic oscillator) does not admit a state with zero
eigenvalue. The lowest state has eigenvalue $i/2$, and plays the
role of the spherical vector in~\eqref{e4},
\begin{equation}
\label{gau}
(E_+-E_-)f=\frac{i}{2}f\ ,\quad f(x)=e^{-\pi x^2}\ .
\end{equation}
Its invariance (up to a phase) under the compact $K$ guarantees that
the theta series~\eqref{e4} depends only on $\tau\in K\backslash G$
(up to a phase). In particular, the $S$ generator, corresponds to
the rotation by an angle $\pi$ inside $K$, and therefore leaves $f$ invariant.
This is the statement that the Gaussian kernel $f$ is 
invariant under Fourier transformation, and lies at the heart
of the automorphic invariance of the theta series~\eqref{e1}. 
The construction holds, in fact,
for any symplectic group $Sp(n,\Zint)$ (with $Sp(1)=Sl(2)$), and
leads to the well known Jacobi-Siegel theta functions,
\begin{equation}
\label{e8}
\theta_{Sp(n,\Zint)}=\sum_{(m^i) \in \Zint^n} e^{i\pi m^i\tau_{ij} m^j}\ .
\end{equation}
This corresponds to the minimal representation
\begin{equation}
\label{e9}
E^{ij}=\frac{i}{2}\, x^i x^j\ ,\quad
E_{ij}=\frac{i}{2}\partial_i \partial_j\ ,\quad
H^i_j=(x^i\partial_j+\partial_j x^i)/2
\end{equation}
of $Sp(n,\Real)$, with algebra
\begin{equation}
[E^{ij},E_{kl}]=\frac{1}{4}\
\Big(
\delta^i_l H^j_k+
\delta^j_l H^i_k+
\delta^i_k H^j_l+
\delta^j_k H^i_l
\Big)\, ,
\end{equation}
acting on the Schwartz space of functions of $n$ variables
$x_i$ (see e.g.,~\cite{Joseph}).  

\subsection{Eisenstein series and spherical vector}
Our second example is the non-holomorphic Eisenstein series 
(see e.g.,~\cite{Terras,Pioline:1998mn})
\begin{equation}
\label{e10}
{\cal E}_s(\tau,\bar\tau)=\sum_{(m,n)\neq (0,0)} 
\left(
\frac{\tau_2}{|m+n \tau|^2}
\right)^s\, ,
\end{equation}
which is a function on upper half plane $U(1)\backslash Sl(2,\Real)$
parameterized by $\tau$ and is 
invariant under the right action of $Sl(2,\Zint)$
given by $\tau\to(a\tau+b)/(c\tau+d)$. This action
can be compensated by a linear one on
the vector $(m,n)$ and the Eisenstein series can therefore 
be rewritten in the symbolic form~\eqref{e4},
where now
$\delta=\sum_{(m,n)\in\Zint^2\backslash(0,0)}\delta(x-m)\ \delta(y-n)$ and
$\rho$ is the linear representation
\begin{equation}
\label{e11}
\rho\begin{pmatrix} a&b  \\ c &d\end{pmatrix}:
\phi(x,y) \to \phi(ax+by, cx+dy)
\end{equation}
corresponding to the infinitesimal generators 
\begin{equation}
\label{e12}
E_+ = x\partial_y\ ,
\quad E_- = y\partial_x \ ,
\quad H = x\partial_x-y\partial_y\ ,
\end{equation}
generating the $Sl(2)$ algebra~\eqref{e7}. The spherical vector
$f(x,y)=(x^2+y^2)^{-s}$ of the representation $\rho$
is clearly invariant under the
maximal compact subgroup $U(1)\subset Sl(2)$
generated by $E_+-E_-$.
In this case, it is not unique (any function of 
$x^2+y^2$ is $U(1)$ invariant) because 
the linear action~\eqref{e11} on functions of
two variables is reducible. An irreducible representation in a single
variable, known as the first principal series, is
obtained by restricting to homogeneous, even functions of
degree $2s$
\begin{equation}
\phi(x,y)=\lambda^{2s} \, \phi(\lambda x,\lambda y)
\end{equation}
and setting $y=1$ (say)
\begin{equation}
\phi(x)\equiv\phi(x,y)\Big|_{y=1}\, .
\end{equation} 
The representation $\rho$ induces an irreducible one
\begin{eqnarray}
\label{fourrep}
&\wt \rho_s\!&\begin{pmatrix} 1 & t\\ 0 & 1 \end{pmatrix}:
\phi(x) \to \phi(x+t) \, ,\label{frog}\\
\label{elephant}
&\wt \rho_s\!&\begin{pmatrix} e^{-t} & 0\\ 0 & e^{t} \end{pmatrix}:
\phi(x) \to e^{-2st}\,\phi(e^{-2t} x ) \, ,\\
&\wt \rho_s\!&\begin{pmatrix} 0 & -1\\ 1 & 0 \end{pmatrix}:
\phi(x) \to x^{-2s} \phi(-1/x)\, .
\label{castle}
\end{eqnarray}
with spherical vector
\begin{equation}
f_s=(x^2+1)^{-s}
\end{equation}

An equivalent representation can be obtained by Fourier transforming
the variable $x$. In terms of the Eisenstein series~\eqref{e10},
this amounts to performing a Poisson resummation on $m$,
\begin{align}
\eis{Sl(2,\Zint)}{2}{s} &= 2\ \zeta(2s)\ \tau_2^{s}
+ \frac{2 \sqrt{\pi}\ \tau_2^{1-s}\ \Gamma(s-1/2)\ \zeta(2s-1)}{\Gamma(s)}
 \nonumber\\
&+\frac{2\pi ^s \sqrt{\tau_2}}{\Gamma(s)}
\sum_{m\neq 0} \sum_{n\neq 0} \left| \frac{m}{n} \right|^{s-1/2}
K_{s-1/2} \left( 2 \pi |mn| \tau_2 \right)
e^{-2\pi i m n \tau_1}\, .
\label{e121}
\end{align}
Using instead the summation variable $N=mn$, this can be rewritten as
\begin{align}
\eis{Sl(2,\Zint)}{2}{s} &= 2\ \zeta(2s)\ \tau_2^{s}
+ \frac{2 \sqrt{\pi}\ \tau_2^{1-s}\ \Gamma(s-1/2)\ \zeta(2s-1)}{\Gamma(s)}
\nonumber\\
&+\frac{2\pi ^s\sqrt{\tau_2}}{\Gamma(s)}
\sum_{N\in\Zint^*} \mu_{s}(N)\; N^{s-1/2} 
K_{s-1/2} \left( 2 \pi \tau_2N \right) 
e^{2\pi i  \tau_1N} \, ,
\label{e122}
\end{align}
where the summation measure of the bulk term 
can be expressed in terms of the number-theoretic quantity 
\begin{equation}
\label{mu33}
\mu_s(N)=\sum_{n| N} n^{-2s+1}\, .
\end{equation}
Indeed, disregarding for now the first two degenerate terms, 
we see that the Eisenstein series can again be written as
in~\eqref{e4}, where the summation measure is 
\begin{equation}
\label{de66}
\delta_s(y)=\sum_{N\in\Zint^*} \mu_{s}(N) \delta(y- N)\, ,
\end{equation}
and the one-dimensional representation $\rho_s$
acting as
\begin{eqnarray}
&\rho_s\!&\begin{pmatrix} 1 & t\\ 0 & 1 \end{pmatrix}:
\phi(y) \to e^{-ity} \phi(y) \, ,\\
\label{e5.6}
&\rho_s\!&\begin{pmatrix} e^{-t} & 0\\ 0 & e^{t} \end{pmatrix}:
\phi(y) \to e^{-2(s-1)t}\,\phi(e^{2t} y ) \, ,
\end{eqnarray}
is generated by
\begin{equation}
\label{e13}
E_+=iy\ , \quad 
E_-=i(y\partial_y +2-2s) \partial_y ,\quad
H=2y\partial_y+2-2s\, .
\end{equation}
Note that this minimal representation has a parameter $s$,
and is distinct from the one in (\ref{e6},\ref{e6.5}). It is,
of course, intertwined with the
representation~(\ref{elephant})-(\ref{castle}) by Fourier transform.
The function 
\begin{equation}
\label{e14}
f_{s}=y^{s-1/2} K_{s-1/2}(y)
\end{equation}
can be easily checked to be annihilated by the compact generator
$K=E_+-E_-=-i(y\pa^2_y+(2-2s) \pa_y-y)$, 
and therefore is a spherical vector of the 
representation~\eqref{e13}. At each value of $s$, it is unique if one 
requires that it
vanishes as $y\to\infty$. 

\subsection{Summation measure, $p$-adic fields and degenerate contributions.}
\label{cowbell}
While the spherical vector can be easily obtained by solving a linear 
differential equation, the distribution $\delta$ invariant
under the discrete subgroup $Sl(2,\Zint)$ appears to be more 
mysterious. In fact, it has a simple interpretation in terms of
$p$-adic number fields, as we now explain.

The simplest instance arises for the $\theta$ series
\eqref{e1} itself which can be rewritten
(at the origin $\tau=i$) as a sum over principal adeles
\begin{equation}
\theta(\tau=i)=\sum_{x\in \Q}\exp(-\pi x^2)\,\prod_{p\;{\rm prime}}
\gamma_p(x)
\end{equation}
where $\gamma_p(x)$ is 1 on the $p$-adic integers
and 0 elsewhere. 
The real spherical vector is the Gaussian and
the function $\gamma_p(x)$ is its $p$-adic analog: 
just like the real Gaussian it is
is invariant under $p$-adic Fourier transform 
(the review~\cite{Brekke:1993}
provides an introduction to $p$-adic numbers and integration theory
for physicists).
Hence $\gamma_p(x)$ is the $p$-adic spherical vector of the representation
\eqref{e5}, and we have thus obtained an ``adelic'' formula for the
unit weight summation measure. 

To take a less trivial case, 
consider the summation measure~\eqref{mu33} appearing in the distribution
$\delta$ in~\eqref{de66}. It can also be rewritten as an infinite product over 
primes,
\begin{equation}
\label{e15}
\sum_N\mu_{s}(N)=\sum_{x\in \Q}\;\prod_{p{\rm\;  prime}} f_p(x)\ ,
\qquad
f_p(x) = \gamma_p(x)\frac{1-p^{-2s+1}|x|_p^{2s-1}}{1-p^{-2s+1}}\, ,
\end{equation}
where $|x|_p$ is the p-adic norm of $N$ (if $N$ is integer, $|N|=p^{-k}$
where $k$ is the largest integer such that $p^k$ divides $N$).
Just as above, $f_p(x)$ can in fact be interpreted as the $p$-adic
spherical vector of the representation~\eqref{e5.6}. To convince oneself
of this fact, one may take the $p$-adic Fourier transform 
of $f_p$, and find 
\begin{equation}
\wt f_p(u)=(1-p^{-2s})^{-1}\,\max(|u|_p,1)^{-2s}\, .
\end{equation}
This is indeed invariant under $u \to -1/u$, and therefore is
a spherical vector for the representation~\eqref{fourrep}\footnote{One may
also check that the product of $\tilde f_p(u)$ over all $p$
reproduces the correct summation measure in the
Eisenstein series~\eqref{e10} upon using the summation variable $u=m/n$.}. 
It is in fact identical to the real spherical vector~\eqref{e10}, upon replacing
the orthogonal real norm $||(x,1)||^2\equiv x^2+1$
by the $p$-adic norm $||(x,1)||_p\equiv \max(|x|_p,1)$.
This suggests that the $p$-adic spherical vector is simply
related to the real spherical vector by changing from orthogonal
to $p$-adic norms and Bessel functions to ``$p$-adic Bessel'' functions.
We shall not pursue this line further here, referring to~\cite{sasha} 
for a rigorous derivation.

Finally, we should say a word about the first two power terms in~\eqref{e122}.
As seen from the above Poisson resummation, these two terms can viewed
as the regulated value of the spherical vector $f(x)$ at $x=0$. 
Unfortunately, we do not know of a direct way to extract them from
$f(x)$ alone; an unsatisfactory method is 
to deduce them by imposing invariance of 
\eqref{e122} under the generator $S$.

\subsection{Generalization to $Sl(n,\Zint)$}
The construction of the minimal representation of $Sl(2,\Real)$
above can be easily generalized to any $Sl(n)$ by starting
with the $Sl(n,\Zint)$ Eisenstein series in the fundamental representation,
\begin{equation}
\label{e16}
\eis{Sl(n,\Zint)}{n}{s}=
\sum_{m^I\in\Zint^n\backslash\{0\}} [m^I g_{IJ} m^J]^{-s}
\end{equation}
and Poisson resumming one integer, $m^1\equiv m$ say. In the language 
of~\cite{Obers:1999um}, this amounts to
the small radius expansion in one direction and we find
\begin{align}
\eis{Sl(n,\Zint)}{n}{s} &=\  2\zeta(2s)R^{-2s}\
+\ \frac{\sqrt{\pi}\ \Gamma(s-1/2)}{R\ \Gamma(s)}\!
\sum_{m^i\in\Zint^{n-1}\backslash\{0\}}\!
[m^i\widehat g_{ij} m^j]^{-s+1/2}  \nonumber\\
&+\ \frac{2\pi ^s}{\Gamma(s)R^{s+1/2}}\!\!\!
\sum_{\stackrel{\scriptstyle m\neq 0}{m_i\in \Zint^{n-1}\backslash\{0\}}} 
\!\!\!\left| \frac{m^2}{ m^i \widehat g_{ij} m^j} 
\right|^{\frac{s-1/2}{2}}\!\!
K_{s-1/2} \left( 2 \pi \frac{|m|}{R} \sqrt{  m^i \widehat g_{ij} m^j} \right)
e^{-2\pi i m m^i A_i}\, .
\label{e161}
\end{align}
We have decomposed the
$n$-dimensional metric $g_{IJ}$ parameterizing
$SO(n,\Real)\backslash Sl(n,\Real)$ into an $n-1$ dimensional metric
$\widehat g_{ij}=g_{ij}-\frac{1}{R^2}\ A_iA_j$, 
the radius of the $n$-th direction $R=g_{11}^{1/2}$ and the off-diagonal
metric $A_i=g_{1i}/g_{11}$.
We now have an $n-1$ dimensional representation of
$Sl(n)$ on $n-1$ variables $x^i$ with $Sl(n-1)$ realized linearly. 
The infinitesimal generators corresponding to positive and negative roots 
are given by
\begin{equation}
\label{e17}
\begin{array}{lcl}
E_+^i=ix^i &\quad& E_{-i}=i(x^j\pa_j+2-2s)\pa_i\, ,\\
E_+^i{}_j=x^i\d_j && E_-^j{}_i=x^j \pa_i\qquad(i>j)\, ,
\end{array}
\end{equation}
with Cartan elements following by commutation.
This is the minimal representation of $Sl(n,\Real)$, generalizing
the $Sl(2,\Real)$ case in~\eqref{e13}. Note that this minimal
representation again has a continuous parameter $s$. For other
groups than $A_n$, the minimal representation will in fact be
unique. For $A_n$, the above representation is unitary
when ${\rm Re}(s)=n/4$.  
The spherical vector is easily read off from~\eqref{e161},
evaluated at the origin $\widehat g_{ij}=g_{ij}=\delta_{ij}$, $R=1$
(rescaling $x^i\rightarrow x^i/(2\pi)$)
\begin{equation}
\label{e18}
f_{A_n,s}= {\cal K}_{s-1/2} \left( \sqrt{ (x^i)^2} \right)
= {\cal K}_{s-1/2}(||(x^1,\ldots,x^{n-1})||)\, ,
\end{equation}
where ${\cal K}_t(x)\equiv x^{-t}K_t(x)$ ($K_t$ is the modified Bessel
function of the second kind) and the Euclidean norm
$||(x_1,x_2,\ldots)||\equiv\sqrt{x_1^2+x_2^2+\cdots}$. This spherical vector
is indeed annihilated by the compact generators following
from~\eqref{e17}.  
The $p$-adic spherical vector in the representation corresponding
to~\eqref{e17} may be obtained
from the summation measure in~\eqref{e161} by the method as outlined
in Section~\ref{cowbell}. The result is
\begin{equation}
f_p(x^1,\ldots,x^{n-1})=\gamma_p(x^1)\cdots\gamma_p(x^{n-1})\,
\frac{1-p^{-s}||(x^1,\ldots,x^{n-1})||_p^s}{1-p^{-s}}\, . 
\end{equation}
Again, this may be obtained from the real spherical vector~\eqref{e15} by
replacing the Euclidean norm by the $p$-adic one
along with  ${\cal K}_s\rightarrow {\cal K}_{p,s}(x)=(1-p^{-s}
x)/(1-p^{-s})$.

\section{Minimal representation for simply laced Lie groups}

The minimal representation we have described for $Sl(n,\Real)$ 
has been generalized in~\cite{KazhdanS} to the case of simply-laced
groups $G(F)$ for arbitrary local field F. In this
Section, we shall review the construction of~\cite{KazhdanS}, and make it
fully explicit.

\subsection{Nilpotent orbit and canonical polarization}

The minimal representation can be understood as the quantization
of the smallest co-adjoint orbit in $\mathcal{G}$.  In order to
construct this minimal orbit, one observes that
all simple Lie algebras have an essentially unique
5-grading (see e.g.,~\cite{Gunaydin:2000xr})
\begin{equation}
\label{e19}
G= G_{-2} \oplus G_{-1} \oplus G_0 \oplus G_1 \oplus G_2
\end{equation}
by the charge under the Cartan generator $H_\omega$ associated
to the highest root $E_\omega$ (for a given choice of Cartan subalgebra
and system of simple roots $\alpha_i$). The spaces $G_{\pm 2}$ have dimension 1
and are generated by the highest and lowest root $E_{\pm \omega}$ respectively.
$G_{1}$ contains only positive roots, and $G_0$ contains all Cartan generators 
as well as the remaining positive roots and the corresponding negative
ones; $G_{-k}$ is obtained from $G_k$ by mapping all positive roots to minus
themselves. The grading~\eqref{e19} can also be obtained by obtained
by branching the adjoint representation of $G$ into the maximal subgroup
$Sl(2)\times H$, where $Sl(2)$ is generated by $(E_{\w},H_\w,E_{-\w})$
and $H$ is the maximal subgroup of $G$ commuting with $Sl(2)$
(explicit decompositions are shown in Table \ref{t5grad} 
for all simply-laced groups) :
\begin{eqnarray}
G &\supset& Sl(2) \times H \nn\\
adj_G&=& (3,1) \oplus (2,R) \oplus (1,adj_H)\\
     &=& 1 \oplus R \oplus [1\oplus adj_H] \oplus R \oplus  1 \nn
\end{eqnarray}
In particular, $G_1$ and $G_{-1}$ transform as a (possibly reducible) 
representation $R$ of $H$, with a symplectic reality condition so that 
$(2,R)$ is real. The set $\Comp H_{\omega} \oplus G_1 \oplus \Comp E_{\omega}$
is the coadjoint orbit of the highest root $E_{\omega}$, namely
the minimal orbit $\mathcal{O}$ we are seeking. Since the highest root
generator $E_{\omega}$ is nilpotent, this is in fact a nilpotent orbit.
As any coadjoint orbit, it carries a standard Kirillov-Kostant symplectic form,
and its restriction to $G_1$ is the symplectic form providing the
reality condition just mentioned.
The nilpotent orbit can also 
be understood as the coset $P\backslash G$ where $P$ is
the parabolic subgroup generated by $G_{-2}\oplus G_{-1}
\oplus (G_0\backslash\left\{H_\omega\right\})$. 
The group $G$ acts on ${\cal O}$
by right multiplication on the coset $P\backslash G$, and therefore
on the functions on ${\cal O}$.

The minimal representation can be obtained by quantizing the orbit
${\cal O}$, {\it i.e.} by replacing functions on the symplectic manifold
${\cal O}$ by operators on the Hilbert space of sections of a line
bundle on a Lagrangian submanifold of ${\cal O}$. In more mundane
terms, we need to choose a polarization, {\it i.e.}
a set of positions and momenta among the coordinates of ${\cal O}$. 
For this, note that, as a consequence of the grading, 
the subspace $G_1\oplus G_2$ forms an
Heisenberg algebra
\begin{equation}
\label{e20}
[E_{\alpha_1},E_{\alpha_2}]= (\alpha_1,\alpha_2) E_{\omega}\, ,\qquad
\alpha_1,\alpha_2\in G_1\, ,
\end{equation}
where $(\cdot,\cdot)$ is the symplectic form. 
A standard polarization 
can be constructing by picking in $G_1$ the simple root $\beta_0$
to which the affine root 
attaches on the extended Dynkin
diagram \footnote{For
$Sl(n)$, the affine root attaches to two roots $\alpha_1$ and
$\alpha_{n-1}$. We choose $\beta_0=\alpha_1$.}. The positive
roots in $G_1$ then split into roots that have an inner product 
$\langle \alpha,\beta_0\rangle$ with $\beta_0$ equal to 1 (we denote them
$\beta_i$), $-1$ (denoted $\gamma_i=\w-\beta_i$), 2 ($\beta_0$ itself),
or 0 (denoted $\gamma_0=\w-\beta_0$). We choose as position operators
$E_{\gamma_0}, E_{\gamma_i}$ and $E_{\omega}$:
\begin{equation}
\label{e21}
E_{\omega}=i y\ , \quad E_{\gamma_i}=i x_i\ \qquad i=0,\dots,d-1
\end{equation}
acting on a space of functions of the variables $y,x_i$. The conjugate
momenta are then represented as derivative operators,
\begin{equation}
E_{\beta_i}= y \partial_i \qquad i=0,\dots,d-1
\end{equation}
The expression for the remaining momentum-like
generator $H_{\w}$ will be determinated
below, but could be obtained at this stage by computing the Kirillov-Kostant
symplectic form on $P\backslash G$.

To summarize our notations at that stage, the 5-grading~\eqref{e19} therefore
corresponds to the decomposition
\begin{equation}
\begin{array}{ccl}
G_{2}&=&\{E_{\omega}\}\, ,\nn\\
G_{1}&=&\{(E_{\beta_i},E_{\gamma_i})\}\nn\\
G_0&=&\{E_{-\a_j},H_{\a_k},E_{\a_j}\}\\
G_{-1}&=&\{(E_{-\beta_i},E_{-\gamma_i})\}\nn\\
G_{-2}&=&\{E_{-\omega}\}
\end{array}
\end{equation}
where $i=0,\ldots,d-1=\dim(R)/2-1$,
$j=1,\ldots,(\dim(H)-\rank(G)+1)/2$
and $H_{\a_k}$ are the Cartan generators of the simple roots
with $k=1,\ldots,\rank(G)$. 

\begin{table}
\begin{equation}
\begin{array}{ccc}
Sl(n) &\supset& Sl(2) \times Sl(n-2) \times \Real^+\\
adj &=& (3,1,0)\oplus [(2,n-2,1)\oplus(2,n-2,-1)] \oplus (1,adj,0) \\
    &=& 1 \oplus 2(n-2) \oplus [1\oplus adj] \oplus 2(n-2) \oplus  1\\[3mm]
SO(2n) &\supset& Sl(2) \times Sl(2) \times SO(2n-4)\\
adj &=& (3,1,1)\oplus(2,2,2n-4)\oplus(1,3,1)\oplus(1,1,adj)\\
    &=& 1 \oplus (2,2n-4) \oplus [1\oplus adj] \oplus (2,2n-4) \oplus 1\\[3mm]
E_6 &\supset& Sl(2) \times Sl(6)\\
78 &=& (3,1) \oplus (2,20) \oplus (1,35)\\
    &=& 1 \oplus 20 \oplus [1\oplus 35] \oplus 20 \oplus 1\\[3mm]
E_7 &\supset& Sl(2) \times SO(6,6)\\
133 &=& (3,1)\oplus (2,32) \oplus (1,66)\\
    &=& 1\oplus 32 \oplus [1\oplus 66]\oplus 32 \oplus 1\\[3mm]
E_8 &\supset& Sl(2) \times E_7\\
248 &=& (3,1)\oplus(2,56)\oplus(1,133)\\
    &=& 1 \oplus 56 \oplus [1\oplus 133] \oplus 56 \oplus 1
\end{array}
\end{equation}
\caption{Five-graded decomposition for simply laced simple groups.
\label{t5grad}}
\end{table}

\subsection{Induced representation and Weyl generators}
Having represented the Heisenberg subalgebra on a space of functions
of $d+1$ variables $(y,x_{i=0,\dots,d-1})$, it remains to extend
this representation to all generators in $G$. This can be
done by unitary induction from the parabolic subgroup
$P$. Rather than taking this approach, we prefer to generate
the missing generators using the unbroken symmetry under $H$
and Weyl generators.

As a first step,
it is useful to note that the choice of polarization
polarization $\Pi$ is invariant under a subalgebra $H_0\subset H$
acting linearly on 
$(x_{i=1,\ldots,d-1})$ while leaving $(y,x_0)$ invariant.
For the $D$ and $E$ groups, $H_0$ is the 
subalgebra generated by the simple roots
which are not attached to $\beta_0$ in the Dynkin diagram of $G$,
whilst for the $A$ series, that by the simple roots attached to neither 
$\beta_0$ nor the root at the other end of the Dynkin diagram.
The subalgebras $H_0$ are listed in Table~\ref{tlin}.

\begin{table}
$$\begin{array}
{c@{\hspace{7mm}}c@{\hspace{7mm}}c@{\hspace{7mm}}c@{\hspace{7mm}}c}
G & \mbox{dim} & H_0 & G_1^* & I_3 \\
Sl(n) & n-1 & Sl(n-3) & [n-3] & 0\\
SO(n,n) & 2n-3 & SO(n-3,n-3) & 1 \oplus [2n-6] & x_1 (\sum x_{2i} x_{2i+1})\\ 
E_6 & 11 & Sl(3)\times Sl(3) & (3,3) & \det   \\
E_7 & 17 & Sl(6) & 15 & \Pf \\
E_8 & 29 & E_6 & 27 & 27^{\otimes_s 3}\vert_{1}
\end{array}$$
\caption{Dimension of minimal representation, linearly realized subgroup
$H_0\subset H\subset G$, representation of $G_1^*$ under $H_0$,
and associated cubic invariant $I_3$.
\label{tlin}}
\end{table}

In order to extend the action of $H_0$ and the Heisenberg subalgebra
to the rest of $G$, we introduce the action of 
two Weyl generators $S$ and $A$. The first, $S$, exchanges 
the momenta $\beta_i$
with the positions $\gamma_i$ for all $i=0,\dots,d-1$ and is therefore achieved
by Fourier transformation in
the Heisenberg coordinates $x_i=0,\dots,d-1$,
\begin{equation}
\label{eS}
(S f)(y,x_0,\dots,x_{d-1})=\int \frac{\prod_{i=0}^{d-1} dp_i}{(2\pi y)^{d/2}}
\, f(y,p_0,\dots,p_d)\,  e^{\frac{i}{y}\sum_{i=0}^{d-1}p_i x_i}
\end{equation}
It also sends all $\alpha_i$ to $-\alpha_i$, while leaving $\omega$
invariant,
\begin{equation}
\label{e22}
S E_{\alpha_i} S^{-1} = E_{-\alpha_i}\ ,\quad
S E_{\omega} S^{-1} = E_{\omega}\ .
\end{equation}
The second generator $A$ is the Weyl reflection with respect to 
the root $\beta_0$.
It maps $\beta_0$ to  minus itself, $\gamma_0$ to $\omega$, 
and all $\beta_i$ to the roots $\alpha_j$ that were not in $H_0$. All 
roots in $H_0$ are invariant under $A$, and so are all 
$\gamma_{i=1,\dots,d-1}$. In order to write the action of~$A$, we
need to introduce an $H_0$-invariant cubic form on $G_1^*$,
\begin{equation}
\label{e23}
I_3=\sum_{i<j<k} c(i,j,k) x_i x_j x_k
\end{equation}
where the sum extends over all $i,j,k=1,\dots,d-1$ such that 
$\beta_i+\beta_j+\beta_k=\beta_0+\omega$. The sign $c(i,j,k)$
is given by~\cite{KazhdanS} 
\begin{equation}
\label{e24}
c(i,j,k)=(-)^{B(\beta_i,\beta_j)+B(\beta_i,\beta_k)+B(\beta_j,\beta_k)
+B(\beta_0,\omega)+1}
\end{equation}
where $B(\alpha,\beta)$ is the adjacency matrix (namely a bilinear form
such that
$\langle\alpha,\beta\rangle=B(\alpha,\beta)+B(\beta,\alpha)$).
The cubic invariant $I_3$ 
in~\eqref{e23} is unique, except for the case of $G=Sl(n)$
where there is none, and is 
listed in the last column 
of Table~\ref{tlin}.
The action of $A$ is given in terms of $I_3$ by
\begin{equation}
\label{eA}
(Af)(y,x_0,x_1,\dots,x_{d-1})
= e^{-\frac{iI_3}{x_0 y}} f(-x_0,y,x_1,\dots,x_{d-1})\, .
\end{equation}
One may check that the generators $A$ and $S$ satisfy the relation
\begin{equation}
\label{e241}
(AS)^3=(SA)^3
\end{equation}
in the Weyl group. In fact, as in the symplectic case where the
relation $(ST)^3$ was equivalent to the invariance of the Gaussian
character $e^{i x^2}$ under Fourier transform, the relation 
\eqref{e241} amounts to the invariance of the cubic character 
$x_0^\alpha (I_3)^\beta e^{i I_3/x_0}$ under Fourier transform over
all $x_{i=0,\dots,d-1}$. This invariance can be easily checked in the
stationary phase approximation\footnote{For $D_4$, the
invariance of the function $(1/|x_0|)e^{ix_1x_2x_3/x_0}$ under
Fourier transform of all 4 variables can be checked explicitly
by performing the (delta function) integrals over $x_1,x_2,x_0,x_3$
in that order.}, and holds exactly for particular values of
the exponents
$\alpha,\beta$~\cite{etingof}. In fact, the minimal representation
yields all
cubic forms $I_3$ such that $e^{i I_3/x_0}$ is invariant~\cite{etingof}.

\subsection{Example: minimal representation of $D_4$}
Using the Weyl generators~\eqref{eS} and~\eqref{eA}, we can now
compute the action of $E_{\alpha}$ in the minimal representation
for all positive and negative roots and in turn obtain the
Cartan generators through $[E_{\alpha},E_{-\alpha}]=\alpha\cdot H$.
As an illustration, we display the $SO(4,4)$ case in detail~\cite{Kazhdan}. 
The data for other groups are tabulated in the Appendix. The Dynkin diagram
of $D_4$ is

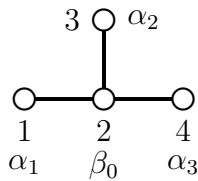
\begin{center}
\begin{picture}(100,35)
\thicklines
\multiput(0,0)(30,0){3}{\circle{8}}
\put(0,-12){\makebox(0,0){1}}\put(0,-25){\makebox(0,0){$\alpha_1$}}
\put(30,-12){\makebox(0,0){2}}\put(30,-25){\makebox(0,0){$\beta_0$}}
\put(60,-12){\makebox(0,0){4}}\put(60,-25){\makebox(0,0){$\alpha_3$}}
\multiput(4,0)(30,0){2}{\line(1,0){22}}
\put(30,4){\line(0,1){22}}
\put(30,30){\circle{8}}
\put(18,30){\makebox(0,0){3}}\put(45,30){\makebox(0,0){$\alpha_2$}}
\end{picture}
\end{center}

\vskip 10mm

\noindent where we have indicated the standard labeling as well as one more
convenient for our purposes; the construction
will be symmetric under permutations of $(\alpha_1,\alpha_2,\alpha_3)$ and
hence under $SO(4,4)$ triality.
The positive roots graded by their  height
along $\beta_0$ are
\begin{equation}
\label{e25}
\begin{array}{ccc}
\alpha_1= &   (1,0,0,0) & =A(\beta_1)\\
\alpha_2= &   (0,0,1,0) & =A(\beta_2)\\
\alpha_3= &   (0,0,0,1) & =A(\beta_3)
\end{array}
\end{equation}
\begin{equation}
\label{e26}
\begin{array}{cc@{\hspace{7mm}}cc}
\beta_0= &   (0,1,0,0)  &  \gamma_0=(1,1,1,1)\\
\beta_1= &   (1,1,0,0)  &  \gamma_1=(0,1,1,1)\\
\beta_2= &   (0,1,1,0)  &  \gamma_2=(1,1,0,1)\\
\beta_3= &   (0,1,0,1)  &  \gamma_3=(1,1,1,0)
\end{array}
\end{equation}
\begin{equation}
\label{e27}
\begin{array}{ccc}
\omega = &   (1,2,1,1)  & =A(\gamma_0)\, .
\end{array}
\end{equation}
We start with the generators
\begin{eqnarray}
\label{e28}
&\begin{array}{c@{\hspace{7mm}}c}
E_{\beta_0}=y\partial_0 & E_{\gamma_0}=i x_0 \\
E_{\beta_1}=y\partial_1 & E_{\gamma_1}=i x_1 \\
E_{\beta_2}=y\partial_2 & E_{\gamma_2}=i x_2 \\
E_{\beta_3}=y\partial_3 & E_{\gamma_3}=i x_3 
\end{array}&\\
\label{e28.5}
&E_{\omega}=i y\, . &
\end{eqnarray}
The cubic form~\eqref{e23} reduces to $I_3=x_1 x_2 x_3$.
The Weyl generator $A$ (acting by conjugation)
yields the generators for the 
remaining simple roots 
$E_{\alpha_i}$ upon which we act with $S$ to obtain 
$E_{-\alpha_i}$,
\begin{equation}
\label{e29}
\begin{array}{c@{\hspace{10mm}}c}
E_{\alpha_1}=- x_{0} \partial_{1} -\frac{i  x_{2} x_{3}}{y} \ ,&
E_{-\alpha_1}= x_{1} \partial_{0} +i  y \pa_{2} \pa_{3} \\
E_{\alpha_2}=- x_{0} \partial_{2} -\frac{i  x_{3} x_{1}}{y} \ ,&
E_{-\alpha_2}= x_{2} \partial_{0} +i  y \pa_{3} \pa_{1} \\
E_{\alpha_3}=- x_{0} \partial_{3} -\frac{i  x_{1} x_{2}}{y} \ ,&
E_{-\alpha_3}= x_{3} \partial_{0} +i  y \pa_{1} \pa_{2} \, .\!\!\!
\end{array}
\end{equation}
A further application of $A$ on $(E_{\beta_0},E_{-\alpha_i})$
yields $(E_{-\beta_0},-E_{-\beta_i})$ upon which $S$ produces the 
$(E_{-\gamma_0},-E_{-\gamma_i})$. Penultimately we may act with $A$ on 
$E_{-\gamma_0}$ to produce the lowest root $E_{-\omega}$,
\begin{eqnarray}
E_{-\beta_{0}}&=&- x_{0} \partial + \frac{i x_{1} x_{2} x_{3}}{y^2}\,  \nn\\
E_{-\beta_{1}}&=& x_{1} \partial + \frac{x_{1}}{y}\,  ( 1+ x_{2} \partial_{2}
+ x_{3} \partial_{3}) - i x_{0} \partial_{2} \partial_{3} \nn\\
\label{e30}
E_{-\gamma_{0}}&=&3i\pa_0+iy\pa\pa_0-y\pa_1\pa_2\pa_3
         +i(x_0\pa_0+x_1\pa_1+x_2\pa_2+x_3\pa_3)\, \pa_0\\
E_{-\gamma_{1}}&=& i y \partial_{1} \partial  + i(2+ x_{0} \partial_{0}
+ x_{1} \partial_{1})\,  \partial_{1}  
- \frac{x_{2} x_{3}}{y}\,  \partial_{0}\nn\\
E_{-\omega}&=&3i\pa + iy\pa^2 +\frac{i}{y} + ix_0 \pa_0\pa+ 
\frac{x_1x_2x_3}{y^2}\, \pa_0 
+ \frac{i}{y}\, (x_1x_2\pa_1\pa_2+x_3x_1\pa_3\pa_1+x_2x_3\pa_2\pa_3)
\nn\\
&& + \ i(x_1\pa_1+x_2\pa_2+x_3\pa_3)\, (\pa+\frac{1}{y})
 +x_0\pa_1\pa_2\pa_3\, ,\nn
\end{eqnarray}
as well as cyclic permutations of (1,2,3), denoting  $\partial\equiv \pa_y$. 
Finally, commutators produce the Cartan generators,
\begin{eqnarray}
H_{\beta_0}&=&- y \partial + x_{0} \partial_{0}\\
H_{\alpha_1}&=&- 1 - x_{0} \partial_{0} + x_{1} \partial_{1} - x_{2} \partial_{2} - x_{3} \partial_{3}\nonumber\label{e31}\\
H_{\alpha_2}&=&- 1 - x_{0} \partial_{0} - x_{1} \partial_{1} + x_{2} \partial_{2} - x_{3} \partial_{3}\nonumber\\
H_{\alpha_3}&=&- 1 - x_{0} \partial_{0} - x_{1} \partial_{1} - x_{2} \partial_{2}
+ x_{3} \partial_{3}\, ,\nonumber
\end{eqnarray}
where $H_{\alpha}\equiv \alpha\cdot H=[E_{\alpha},E_{-\alpha}]$ 
is the Cartan generator along the simple root
$\alpha$. Note that the Cartan generator corresponding to the highest
root $\omega$ has a simple form,
\begin{equation}
\label{e32}
H_{\omega}=[E_\omega,E_{-\omega}]=-3 -2y\pa -x_{0} \partial_{0}-
x_{1} \partial_{1} - x_{2} \partial_{2}- x_{3} \partial_{3}
\end{equation}
and therefore acts by a uniform rescaling on all $x_i$, and 
a double rescaling on $y$. This agrees with the fact that $H_{\omega}$
is the grading operator in~\eqref{e19}. The form of this expression 
holds therefore for all groups (save for the non-universal
constant term $-3$ above). This is also true of  the expression 
for $H_{\beta_0}$.
The generators for the positive and negative
simple roots and Cartan elements for all simply-laced 
groups, computed following the same procedure, are given in the Appendix. 
All other roots can be obtained by commuting
\begin{equation}
[E_\alpha,E_\beta]=
\left\{
\begin{array}{cl}
\pm \, E_{\alpha+\beta}
&\mbox{ if $\alpha+\beta$ is a root} \\
0& \mbox{ otherwise}\, .
\end{array}\right.
\end{equation}
Finally, to specify our conventions for positive and negative roots,
we record that the quadratic Casimir operator for $D_4$ is
\begin{equation}
C=\sum_i H_{i}C^{ij}H_{j}
-\sum_{k}\E_{\a_k}+\sum_{l}(\E_{\beta_l}+\E_{\gamma_l})
-\E_{\beta_0}-\E_{\gamma_0}-\E_\omega\, ,
\end{equation}
The same formula may be applied to other groups as well.
Here $C^{ij}$ is the inverse Cartan matrix and $\E_\a\equiv
E_\a E_{-\a}+E_{-\a}E_{\a}$. Evaluated on the minimal representation
above, we have $C=-8$, in agreement with irreducibility. 
Note however that in contrast to ordinary representations, the center of 
the minimal representation (Joseph's ideal) is much larger:
e.g, any quadratic polynomial in the Cartan generators $H_i$ can be 
supplemented with a linear combination of $\E_\alpha$ operators 
to make a scalar element.

To summarize, we have obtained a unitary irreducible representation
of any simply-laced split group $G$ by quantizing the action of
$G$ on its minimal nilpotent orbit (the classical limit can be
obtained from our formulae for the generators
by replacing the derivative operators $i(\pa,\pa_i)$ by 
momenta $(p,p_i)$ conjugate to the coordinates $(y,x_i)$
and dropping the ``normal ordering'' terms; this yields
the Hamiltonians for the generators of $G$ on the nilpotent
orbit with symplectic form $dp\wedge dy+\sum dp_i\wedge dx_i$). 
Choosing one of the generators as the Hamiltonian gives a dynamical
system with a spectrum generating symmetry $G$. This generalizes
the $A_1$ case corresponding to conformally invariant quantum
mechanics \cite{fubini}.

\section{Spherical vectors for $D_n$ and $E_{6,7,8}$ Lie groups}

With the explicit minimal representation 
for all simply-laced groups at hand, 
we focus our attention on the
spherical vector; a function $f(y,x_i)$ annihilated by
all compact generators in $G$. This is our main result
and is a central building block for the construction of
theta series for all groups.

\subsection{From symplectic to orthogonal}
One way of obtaining the symplectic vector is to solve the
differential equation $(E_{\alpha}\pm E_{-\alpha})f=0$
for all roots $\alpha$ (the sign is chosen so that the generator
is compact). It is sufficient to solve these equations
for $\alpha$ a simple root only, since all other equations can be obtained
by commutation. This still sounds like a formidable task, even though
we shall in fact be able to carry it out later on for exceptional groups.
For now however, we would like to take an alternate approach,
well suited to orthogonal groups. The spherical vector we shall obtain
will turn out to generalize quite simply to exceptional groups as well.

The main observation is that there is a maximal embedding of
$SO(n,n,\Real) \times Sl(2,\Real)$ in $Sp(2n,\Real)$. The minimal
representation of $Sp(2n,\Real)$ has dimension $2n$, and is also a 
representation of $SO(n,n,\Real)$, albeit reducible. By considering
functions invariant under $Sl(2,\Real)$ however, we can reduce
it to a $2n-3$ dimensional representation, which is the dimension
of the minimal representation. In this way we thus obtain a
representation equivalent to the one described in Section~3.
In order to obtain the spherical vector in that representation,
we just need to integrate over the second factor in the decomposition
\begin{equation}
\label{e33}
\frac{SO(n,n,\Real)}{SO(n)\times SO(n)}
\times \frac{Sl(2,\Real)}{U(1)}
\subset
\frac{Sp(2n,\Real)}{U(2n,\Real)}
\end{equation}
to get a function on the first space.

This procedure is familiar to string theorists
since it gives precisely the one-loop result
for half-BPS amplitudes. Indeed, the partition function of
the worldsheet winding modes on a torus $T^n$ is a theta series for the
symplectic group $Sp(2n,\Real)$, restricted to the subspace~\eqref{e33}
of the moduli space. It can be written in a form which makes the
modular symmetry $Sl(2,\Zint)$ manifest,
\begin{equation}
\label{e34}
\theta_{Sp}=V\,
\sum_{m^i,n^i} \exp\Big(-\pi\ \frac{(m^i+n^i\tau)\, g_{ij}
(m^j+n^j\bar\tau)}{\tau_2}+2\pi i m^iB_{ij}n^j\Big)\, .\,
\end{equation}
where we recognize a sum weighted by the Polyakov action for classical
toroidal strings winding around $T^n$ with volume $V=\surd\det g_{ij}$
via $X^i(\sigma_1,\sigma_2)=m^i\sigma_1+n^i\sigma_2$;
or with manifest $SO(n,n,\Zint)$ target space symmetry,
\begin{equation}
\theta_{Sp}=\tau_2^{n/2}\,
\sum_{m_i,n^i}\ \exp
\left(
-\pi\tau_2\,\Big[ (m_i+B_{ik}n^k)\, 
g^{ij}(m_j+B_{jl}n^l)+n^ig_{ij}n^j\Big]+2\pi i \tau_1 m_in^i\right)\, .
\label{e35}
\end{equation}
In this form, we recognize the contribution of states with
momentum $m_i$ and winding $n^i$ in the Schwinger representation,
with a BPS constraint $m_i n^i=0$. The two representations 
are related by Poisson resummation over all Kaluza--Klein modes
$m^i\leftrightarrow m_i$.
The one-loop amplitude is obtained by
integrating this theta series over the fundamental domain of the 
upper half-plane $U(1)\backslash Sl(2)$ parameterized by the
worldsheet modulus $\tau$:
\begin{equation}
\label{e36}
\theta_{SO(n,n)}(g_{ij},B_{ij})=2\pi\,\int_{\F} \frac{d^2\tau}{\tau_2^2}\, 
\theta_{Sp}(\tau,\bar\tau;g_{ij},B_{ij})\, .
\end{equation}
The result is an automorphic form under the T-duality group
$SO(n,n,\Zint)$. Its expansion at large volume using the methods
described, e.g., in~\cite{Obers:1999um}, reads
\begin{equation}
\label{e37}
\theta_{SO(n,n)}=
\frac{2\pi^2}{3}V
+2V \!\sum_{m^i\neq 0}\frac{1}{ m^{i}g_{ij}m^{j}}
+4\pi V \!\!\!\!
\sum_{(m^{i},n^i)/Sl(2)} \!\!\!\!
\frac{e^{-2\pi\,\sqrt{(m^{ij})^2}+2\pi i m^{ij}B_{ij}}}{\sqrt{(m^{ij})^2}}\ ,
\end{equation}
and exhibits a sum of power-suppressed contributions, together with
worldsheet instantons. The double sum runs over integer vectors $(m^i,n^i)$
modulo the linear action of $Sl(2,\Zint)$. The worldsheet instantons however
depend only on the $Sl(2)$ invariant combination $m^{ij}=m^in^j-m^j n^i$
(with $(m^{ij})^2\equiv\frac{1}{2!}\,m^{ij}g_{ik}g_{jl}m^{kl}$),
so they can be rewritten as
\begin{equation}
\label{e38}
\sum_{m^{ij} \mbox{rank 2}} \mu(m^{ij})\, 
\frac{e^{-2\pi\sqrt{(m^{ij})^2}+2\pi i m^{ij}B_{ij}}}{\sqrt{(m^{ij})^2}}
\end{equation}
where the measure factor $\mu(m^{ij})=\sum_{n|m^{ij}} n$ accounts for
the Jacobian factor between variables $(m^i,n^i)$ and $m^{ij}$. 
We thus have a representation
of $SO(n,n,\Real)$ on a space of rank 2 antisymmetric matrices $m^{ij}$. The 
dimension of this space is precisely $2n-3$ and ought therefore 
be equivalent to the minimal representation described in Section~3.
We can also read off the real spherical vector immediately
by going to the origin $(g_{ij}=\delta_{ij},B_{ij}=0)$ of the moduli
space,
\begin{equation}
\label{e39}
\wt f_{D_n}=\frac{e^{-2\pi \sqrt{(m^{ij})^2}}}{\sqrt{(m^{ij})^2}}
\end{equation}
The $p$-adic spherical vector can also be extracted from the
summation measure $\mu(m^{ij})$ in the same way as in~\eqref{e15},
and reads
\begin{equation}
\label{e39p}
\wt f_p=\gamma_p(m_{ij}) \frac{1-p \,||(m_{ij})||_p}{1-p}\, .
\end{equation}

\subsubsection*{$D_4$ spherical vector in the standard minimal representation.}

Having found the spherical vector in this ``string inspired''
representation, we now would like to map it to the standard
minimal representation,
with the aim of generalizing it to exceptional groups. For this
we need to find the linear operator that intertwines between
the two representations and let it act on the spherical vector~\eqref{e39}.
For simplicity, we will describe the $SO(4,4)$ case only, since the
method generalizes easily to higher $n$. In this case, the constraint
that $m^{ij}$ of the ``string-inspired'' representation has rank 2 is
\begin{equation}
\label{e40}
\epsilon_{ijkl} m^{ij} m^{kl} = 0\, ,
\end{equation}
which describes a 
quadratic cone in $\Real^6$. 

Firstly, consider the operator acting by multiplication by
$m_{ij}$ (corresponding
to shifting $B_{ij}$ by a constant) in the representation~\eqref{e37}.
These shifts make a 6-dimensional Abelian subalgebra of the Borel
subgroup of $SO(4,4)$ ({\it i.e.}, the group generated by the positive roots).
We can identify six commuting generators 
by choosing those for roots with height one in the direction of, for example, 
$\alpha_3$, namely 
$(E_{\a_3},E_{\b_3},E_{\g_1},E_{\g_2},E_{\g_0},E_\omega)$.
Since these operators commute, we can diagonalize them simultaneously
and we call their eigenvalues
$i(m^{43},m^{24},m^{14},$ $m^{23},m^{13},m^{12})$.
Using the expressions in~\eqref{e28}-\eqref{e30}
for the generators, we find a
common eigenstate
\begin{equation}
\label{e41}
\psi_{m^{ij}}=\delta(y-m^{12})\,\delta(x_0-m^{13})\,\delta(x_1-m^{14})\,
\delta(x_2-m^{23})\,e^{\frac{im^{24}\,x_3}{m^{12}}}\ ,
\end{equation}
but only if the eigenvalues are related by
\begin{equation}
\label{e42}
m^{43}=-\frac{m^{14}m^{23}}{m^{12}}-\frac{m^{13}m^{24}}{m^{12}}\, .
\end{equation}
This is the same as~\eqref{e40}, providing the rationale for our
identification. 
Therefore the two representations are intertwined by Fourier transformation
in a single variable $x_3$,
\begin{eqnarray}
\wt f(m^{ij})&=&\int dy dx_0 d^3x\,\psi_{m^{ij}}(y,x_0,x)\,f(y,x_0,x)\nn\\
&=&\int dx_3
\exp(im^{24}\,x_3/m^{12})\,f(m^{12},m^{13},m^{14},m^{23},x_3)\, .
\label{e43}
\end{eqnarray}
where $x$ stands for $(x_1,x_2,x_3)$.
Conversely, we have
\begin{eqnarray}
f(y,x_0,x)&=&\int \frac{dm^{24}}{y}\,
e^{-\frac{2\pi im^{24}\,x_3}{y}}
\wt f\left(y,x_0,x_1,x_2,m_{24},\frac{x_1 x_2 + x_0 m^{24}}{y}\right) 
\label{e44}\\
&=&\int dm^{24} dm^{43}
e^{-\frac{2\pi i m^{24}\,x_3}{y}}\delta(x_1 x_2+x_0 m^{24}+y m^{43})
\wt f(y,x_0,x_1,x_2,m^{24},m^{43})\, , \nn
\end{eqnarray}
where 
$\wt f_{m^{ij}}\equiv\wt f(m^{12},m^{13},m^{14},$ $m^{23},m^{24},m^{43})$.

To see how the kernel~\eqref{e39} translates into 
the standard minimal representation we must
compute the Fourier transform~\eqref{e44}. For that purpose,
it is convenient to take the integral representation
\begin{equation}
\label{e45}
\wt f=\frac{e^{-2\pi\sqrt{(m^{ij})^2}}}{\sqrt{(m^{ij})^2}}
=\int_0^{+\infty} \frac{dt}{t^{1/2}}\,
\exp\left(-\pi/t-\pi t (m^{ij})^2\right)\, ,
\end{equation}
along with the standard one for the Dirac delta function of the constraint.
Hence, the action of the intertwining operator on the string-inspired
spherical vector may be written as
\begin{equation}
\label{e46}
f=\int_0^\infty\!\!\! \frac{dt}{t^{1/2}} \int_{-\infty}^{+\infty}
\!\!\!\!\!\! d\theta\,
dm^{24} dm^{43}
e^{-\frac{\pi}{t}-\pi t (m^{ij})^2
-2\pi i \theta(x_1 x_2+x_0 m^{24}+y m^{43})
-2\pi i  \frac{m^{24}\,x_3}{y}}\,\Big
|_{{}^{\sss m^{12}=y,}_{\sss m^{14}=x_1,}
{}^{\sss m^{13}=x_0,}_{\sss m^{23}=x_2}}\, .
\end{equation}
The integrals over $m^{24},m^{43}$ are Gaussian and yield
\begin{equation}
\label{e47}
f=\int \frac{dt}{t^{3/2}} \int_{-\infty}^{+\infty} \!\!d\theta\, 
e^{-\pi t (y^2+x_0^2+x_1^2+x_2^2)
-\frac{\pi}{t} \left(1+ (\theta y)^2+(\theta x_0-\frac{x_3}{y})^2 \right)
-2\pi i \theta x_1 x_2}\, .
\end{equation}
The integral over $\theta$ is again Gaussian,
and the $t$ integral is of Bessel type so all integrals can be computed
explicitly. The saddle point yields a classical action at
\begin{equation}
\label{e48}
S=2\pi\, \frac{\sqrt{(y^2+x_0^2+x_1^2)(y^2+x_0^2+x_2^2)(y^2+x_0^2+x_3^2)}}
{y^2+x_0^2} - 2\pi i \frac{x_0 x_1 x_2 x_3}{y (y^2 + x_0^2)}\, .
\end{equation} 
Taking into account the measure factor, we find that 
in the standard representation, the kernel becomes
(rescaling all variables $(y,x_0,x_1,x_2,x_3)$ by $1/(2\pi)$)
\begin{equation}
\label{e49}
f_{D_4}=\frac{4\pi}{\sqrt{y^2+x_0^2}}\, K_0 \left( 
\frac{\sqrt{(y^2+x_0^2+x_1^2)(y^2+x_0^2+x_2^2)(y^2+x_0^2+x_3^2)}}
{y^2+x_0^2} \right)
e^{ - i\, \frac{x_0 x_1 x_2 x_3}{y (y^2 + x_0^2)}}\, .
\end{equation}
This expression is the prototype of the spherical vectors that we
will obtain later on, and therefore deserves several comments:
\begin{itemize}

\item[(i)] It is invariant under permutations of 
$(x_1,x_2,x_3)$, {\it i.e.}, under $SO(4,4)$ triality which
was manifest in the standard representation but not at all
in the string-inspired one. In fact, 
on the basis of Heterotic/type II duality, it was found
that
the one-loop string amplitude~\eqref{e36} for $n=4$ would have
to be invariant under triality ~\cite{Kiritsis:2000zi}
(see also~\cite{Nahm:2001ps} for a related observation). 
Therefore our triality
invariant result 
gives strong support to non-perturbative
Heterotic/type II duality.

\item[(ii)]
The spherical vector~\eqref{e49} could also have been derived
by solving the differential equations for $K$-invariance. As we will
show for exceptional cases later, the system of PDE's reduces to
a single differential equation for a single function of the variable
$S_1$, 
\begin{equation}
\label{e511}
S_1=\frac{\sqrt{(y^2+x_0^2+x_1^2)(y^2+x_0^2+x_2^2)(y^2+x_0^2+x_3^2)}}
{y^2+x_0^2}\ .
\end{equation} 
This is equation is a linear second order differential equation 
of Bessel type, for which~\eqref{e49} is the 
only solution with exponential decrease
at infinity ({\it i.e.}, $S_1\to \infty$). The same phenomenon will also hold
for exceptional groups, except that the variable $S_1$ will be a more 
complicated function of the coordinates $(y,x_i)$ 
(but reducing to the same form~\eqref{e511} for particular
configurations of the variables $x_i$), and that the order of the 
Bessel function will be different.

\item[(iii)] The phase $\exp(-iS_2)$, where $S_2$ is the imaginary part of the 
classical action
\begin{equation}
\label{e495}
S_2=\frac{x_0I_3}{y(y^2+x_0^2)}\, ,
\end{equation}
is precisely such that the spherical vector
is invariant under the Weyl generator A in~\eqref{eA}. Indeed,
this follows from the trivial identity
\begin{equation}
\label{e50}
\frac{I_3}{x_0 y} -\frac{y I_3}{x_0(x_0^2+y^2)}
=\frac{x_0 I_3}{y(x_0^2+y^2) }\, .
\end{equation}
Defining 
\begin{equation}
\label{e51}
f(y,x_i)\equiv g(y,x_i)\, e^{ -i \frac{x_0 I_3 }{y (y^2 + x_0^2)}}
\end{equation}
we see that the invariance under $A$ requires $g$ to be 
symmetric under $(y,x_0)\to(-x_0,y)$. In fact, the invariance
under the compact generator $E_{\beta_0}+E_{-\beta_0}$ requires
$g$ to depend on $(y,x_0)$ through $y^2+x_0^2$ only since it acts on
the function $g(y,x_i)$ as the rotation operator 
$y\partial_0-x_0\partial$. 
This will hold for all simply-laced groups.

\item[(iv)] In the limit $(y,x_0)\to 0$, due to the asymptotic
behavior $K_s(y\to \infty)\sim e^{-y}\sqrt{\pi/(2y)}$, the spherical vector
takes a much simpler form
\begin{equation}
f_{D4}\sim \frac{1}{\sqrt{|x_1x_2x_3|}}
e^{-\frac{|x_1x_2x_3|}{yz}} \quad \mbox{or} \quad
\frac{1}{\sqrt{|x_1x_2x_3|}}
e^{-\frac{|x_1x_2x_3|}{y\bar z}}
\end{equation}
depending of the sign of $x_1x_2x_3$, where $z=y+ix_0$. We recognize
the same kernel as in the definition of the Weyl generator $A$
in~\eqref{eA}. The spherical vector~\eqref{e49} 
can therefore be thought of as 
a Fourier-invariant non-linear (physicists would say
``Born-Infeld'') completion of the Fourier invariant
kernel in~\eqref{eA}.

\item[(v)] The result for the spherical vector~\eqref{e49} may be rewritten
more compactly in terms of the Euclidean norm 
$||(x_1,x_2,\ldots)||\equiv\sqrt{x_1^2+x_2^2+\cdots}$ as
\begin{equation}
\label{fgrad}
f=\frac{1}{R}\, K_0\Big(||(X,\nabla_{\!\!X}\Big(\frac{I_3}{R}\Big))||\Big)\, 
\exp(-i\, \frac{x_0I_3}{yR^2})\, ,
\end{equation}
where $R=||(y,x_0)||$, $X\equiv(y,x_0,x_1,x_2,x_3)$ and 
$\nabla_{\!\!X}(I_3/R)$ denotes the gradient of $I_3/R$ with respect to 
the $X$ coordinates.

\item[(vi)]  The $p$-adic spherical vector in the string inspired
representation can be read off from the large volume expansion~\eqref{e37},
\begin{equation}
f_p(m^{ij},m\wedge m=0)
=\Big(\prod_{i<j}\gamma_p(m^{ij})\Big)\,\frac{1-p\,||(m^{ij})||_p}{1-p}\, .
\end{equation}
The corresponding spherical vector in the triality invariant representation
can be obtained via intertwining by $p$-adic Fourier transform. We
leave the details of this computation to \cite{sasha}, and simply
mention that it takes the same form as \eqref{fgrad}, upon replacing
the Euclidean norm with the $p$-adic norm $||(x_1,x_2,\ldots)||
=\max(|x_1|_p,|x_2|_p,\ldots)$, and $K_0$ by a simple
algebraic function.

\end{itemize}

\subsubsection*{$D_n$ spherical vector in the standard minimal representation.}
Before moving on to exceptional groups, let us note that the same
manipulation can be performed for higher $SO(n,n)$ groups. 
In the string representation there are $n(n-1)/2$ variables $m^{ij}$
subject to constraints 
\begin{equation}
\epsilon_{i_1\ldots i_{d-4}ijkl}m^{ij}m^{kl}=0
\Leftrightarrow
m^{[ij}m^{kl]}=0\, .
\end{equation}
Of these, only $(n-2)(n-3)/2$ namely $m^{1[2}m^{kl]}=0$ (say), are independent,
so the dimension of the minimal representation is
\begin{equation}
\frac{n(n-1)}{2}-\frac{(n-2)(n-3)}{2}=2n-3
\end{equation}
as given in Table~\ref{tlin}.

The intertwining
operator is a Fourier transform on $n-3$ variables, and its action
on the spherical vector~\eqref{e39} can be computed using the same
manipulations as before. We quote
\begin{equation}
\label{fdn}
f_{D_n}=\left(\frac{y^2+x_0^2+x_1^2}
{(y^2+x_0^2)^2+(y^2+x_0^2)P+Q^2}
\right)^{\frac{n-4}{4}}\,
\frac{ K_{\frac{n-4}{2}}(S_1)e^{-iS_2}}{\sqrt{y^2+x_0^2}}\, ,
\end{equation}
where
\begin{equation}
\label{eSdn}
S_1=\frac{\sqrt{(y^2+x_0^2)^3+(y^2+x_0^2)^2 I_2
+(y^2+x_0^2)\, (I_2^2-I_4)/2 +(I_3)^2}}{y^2+x_0^2}\, ,
\end{equation}
\begin{equation}
S_2=\frac{x_0 I_3}{y (y^2 + x_0^2)}
\end{equation}
and
\begin{equation*}
P=\sum_{j=2}^{2n-5} x_j^2\ ,\quad
Q=\sum_{i=1}^{n-3} (-)^{i+1}\, x_{2i} x_{2i+1}\ ,
\end{equation*}
\begin{equation}
I_2=x_1^2+P\ ,\quad
I_3=x_1 Q\ ,\quad
I_4=x_1^4+ P^2 -2 Q^2\ .
\end{equation}

In contrast to the $n=4$ case, for $n>4$ the spherical vector
must be invariant under the maximal compact subgroup 
$K_0=SO(n-3)\times SO(n-3)$ of the linearly realized
$H_0=SO(n-3,n-3,\Real)$. This is indeed the case of our result,
since $P$ and $Q$ are the $K_0$-invariant square norms of
the $SO(n-3,n-3)$-vector $(x_2,\dots,x_{2n-5})$ ($Q$ is even $H_0$ invariant). 
Using this symmetry we can choose
all $x_{i>3}$ to vanish. In this case the classical action, 
$S=S_1+iS_2$ reduces to the
$D_4$ case~(\ref{e511},\ref{e495}). 
As for $D_4$, we can express the $D_n$ spherical vector 
more compactly as 
\begin{equation}
f_{D_n}=\frac{1}{R}\Big(\frac{
||(y,x_0,x_1)||}{R}\Big)^{n-4}\, 
{\cal K}_{\frac{n-4}{2}}\Big(||(X,\nabla_{\! X}
\Big(\frac{I_3}{R}\Big)||\Big)\, \exp\Big(-i\frac{x_0I_3}{yR^2}\Big)\, .
\label{norms}
\end{equation}
Here ${\cal K}_t(x)\equiv x^{-t}K_t(x)$, $X\equiv(y,x_0,x_1,\ldots,x_{2n-5})$ 
and $R\equiv||(y,x_0)||$. The form~\eqref{eSdn} of 
the argument of the Bessel function 
 in terms of the three $K_0$-invariants $I_2,I_3,I_4$,
will apply to the exceptional groups as well as will as
the overall form~\eqref{norms}.

As in the $D_4$ case, the $p$-adic spherical vector in the string
inspired representation can be read off from the large volume expansion of the
symplectic theta series. The spherical vector  in the ``standard''
minimal representation could therefore be obtained by
$p$-adic Fourier transform.

\subsection{$E_6$}
In the case of exceptional groups, we unfortunately do not have
a string-inspired representation which we could use to obtain 
the spherical vector. In fact, it is the other way around, since
we are aiming at a ``membrane-inspired'' representation 
for exceptional theta series! Our only remaining line of attack is
therefore to find an explicit solution of the differential
equations $(E_{\a}\pm E_{-\a})f=0$ determining the spherical vector.

For this, let us recall that (i) once the phase factor in 
\eqref{e51} is factored out, the dependence of $f$ on $(y,x_0)$
is through $(y^2+x_0^2)$ only, and (ii) that the spherical vector
has to be invariant under the maximal compact subgroup $K_0$ of $H_0$,
which is linearly realized on $(x_1,\dots,x_d)$. Our first task, therefore,
is to determine the invariants of  $(x_1,\dots,x_d)$
under $K_0$.

In the $E_6$ case, from Table~\ref{tlin} the variables $(x_1,\dots,x_9)$
transform in a $(3,3)$ representation of $H_0=Sl(3)\times Sl(3)$. 
Using the $K_0$ transformations implied by the
explicit expressions for the roots given in the Appendix,
we can assign the 9 variables to a $3\times 3$ matrix
\begin{equation}
\label{e52}
Z=\begin{pmatrix}
x_1&x_3&x_6\\
x_2&x_5&x_9\\
x_4&x_7&x_8
\end{pmatrix}\, ,
\end{equation}
on which $Sl(3)\times Sl(3)$ act linearly by left and right multiplication
respectively. An independent set of invariants
under the maximal compact subgroup
$K_0=SO(3)\times SO(3)$ is given by the quadratic, cubic and quartic
combinations
\begin{equation}
\label{e53}
I_2 = \Tr(Z^tZ) \ ,\qquad I_3=-\det(Z)\ ,\qquad I_4=\Tr(Z^tZZ^tZ)\, ,
\end{equation}
In fact, $I_3$ is our familiar cubic form, invariant under the whole
of $H_0$ and not only its maximal compact subgroup. Note also that
higher traces are algebraically related to the ones above.

Now, given that the spherical vector has to be invariant under $K_0$,
we can work in a frame where $Z$ is diagonal keeping 
$(x_1,x_5,x_8)$ as the
only non-vanishing entries. The invariants then reduce to
\begin{equation}
\label{e54}
I_2=x_1^2+x_{5}^2+x_{8}^2\ ,\quad
I_3=-x_1 x_{5} x_{8}\ ,\quad
I_4=x_1^4+x_{5}^4+x_{8}^4\ .
\end{equation}
Let us now consider the equation $(E_{\beta_1}-E_{-\beta_1})f=0$.
The negative root generator $E_{-\beta_1}$ can be obtained by commuting
the negative roots given in the appendix, and reads
\begin{eqnarray}
E_{-\beta_1}&=&x_1\partial+ix_0(\pa_5\pa_8-\pa_7\pa_9)+\frac{2}{y} x_1
+\frac{1}{y}\left(x_1(x_5\pa_5+x_7\pa_7+x_8\pa_8+x_9\pa_9) \right.\nn\\
&&\left.
-x_2x_3\pa_5-x_2x_6\pa_9-x_3x_4\pa_7-x_4x_6\pa_8\right)\ . 
\end{eqnarray}
Using the ansatz~\eqref{e51} and setting all $x_{i=0,\dots,9}$
but $(x_1,x_5,x_8)$ to
zero at the end, we get a first order differential equation
\begin{equation}
\label{e55}
x_1(x_5 \pa_5 +x_8\partial_8-2)\, g +y^2(\pa_1+2 x_1\pa_y)\,  g=0\, ,
\end{equation}
which is solved by $g(y^2,x_1,x_5,x_8)=\frac{1}{y^2}
h(y^2+x_1^2,\frac{x_5}{y},\frac{x_8}{y})$.
{}Demanding invariance under the compact generators of
$\beta_5$ and $\beta_8$ requires the same equation to hold
for permutations of $x_1,x_5,x_8$ so the only possibility is 
\begin{equation}
g(y^2,x_1,x_5,x_8)=\frac{1}{y^2}\, 
h\left(\frac{\sqrt{ (y^2+x_1^2)(y^2+x_5^2)(y^2+x_8^2)}}{y^2}\right)\, .
\end{equation} 
The 
argument of $h$ is easily recognizable as the universal form $S_1$ 
in~\eqref{e511} 
and we can 
restore the dependence on all variables using $K_0$ invariance;
first in our particular frame we write
\begin{eqnarray}
S_1&=&\frac{\sqrt{(y^2+x_1^2)(y^2+x_5^2)(y^2+x_8^2)}}{y^2} \nn\\
&=&\frac{\sqrt{y^6+y^4(x_1^2+x_5^2+x_8^2)+y^2(x_1^4+x_5^4+x_8^4)+
(x_1x_5x_8)^6}}
{y^2}\, .
\end{eqnarray}
Then using the 
relations~\eqref{e54} for the  invariants $I_2$, $I_3$ and $I_4$,
and recalling that the dependence on $(y,x_0)$ is through the
norm $y^2+x_0^2$, we find that $h$ has to depend on the 
coordinates $(y,x_0,\dots,x_ 9)$ through the combination
\begin{eqnarray}
\label{e562}
S_1=
\frac{
\sqrt{(y^2+x_0^2)^3 + (y^2+x_0^2)^2 I_2 + (y^2+x_0^2)(I_2^2 - I_4)/2 + I_3^2}}
{(y^2+x_0^2)}\, .\label{e56}
\end{eqnarray}
In fact, this expression can be rewritten in a much more concise way as
\begin{equation}
\label{e563}
S_1 = \frac{\sqrt{\det( Z Z^t + |z|^2 \mathbb{I}_3)}}{|z|^2}\, ,
\end{equation}
where $z=y+i x_0$ and $\mathbb{I}_3$ is the $3\times 3$ identity matrix.
This expression is manifestly invariant under 
$SO(3)\times SO(3)\times SO(2) \subset K$.

Finally, the $(E_{\alpha_3}-E_{-\alpha_3})f=0$ equation reduces
to 
\begin{equation}
h''+\frac{2}{S_1} h'-h=0\, .
\end{equation} 
This is a Bessel-type equation, solved
by $h=K_{1/2}(S_1)/\sqrt{S_1}\propto e^{-S_1}/S_1$.
Altogether, we have thus found an explicit expression for the $E_6$
spherical vector in the minimal representation,
\begin{equation}
\label{efE6}
f_{E_6}=\frac{e^{-S_1-i\frac{x_0 I_3}{y(x_0^2 + y^2)}}}{(y^2+x_0^2)S_1}
\propto\frac{K_{1/2}(S_1)e^{-i\frac{x_0 I_3}{y(x_0^2 + y^2)}}}
{(y^2+x_0^2)\,\sqrt{S_1}}
\, .
\end{equation}
As in the $D4$ case,
this expression simplifies greatly in the limit $|z| \to 0$,
\begin{equation}
\label{efE6l}
f_{E_6}\sim \frac{e^{-\frac{|\det Z|}{yz}}}{|\det Z|}
\end{equation}
or its complex conjugate, depending on the sign of $\det(Z)$.

While this spherical vector has been constructed in the standard
representation presented in Section~3, other choices of polarization
can be relevant in certain applications, and yield different expressions
for the spherical vector. In the $E_6$ case, there is another interesting
polarization, where the linearly realized group is $Sl(5)$ rather than
$Sl(3)\times Sl(3)$. By looking at the root lattice displayed in the
appendix, it is easy to see that this representation can be reached
by performing a Fourier transform on the coordinates ($x_6,x_9,x_8)$.
This breaks one of the $Sl(3)$ factors, but the other unbroken
factor gets enlarged to an $Sl(5)$, under which the 10 new coordinates
transform as an antisymmetric matrix
\begin{equation}
X=\begin{pmatrix} 0 & -p_8&p_9&x_1&x_3\\
& 0  & -p_6&x_2&x_5\\
&    & 0 & x_4 & x_7\\
&a/s    &   & 0   &x_0\\
&    &   &     &0
\end{pmatrix}
\end{equation}
where $(p_6,p_8,p_9)$ are the momenta conjugate to $(x_6,x_8,x_9)$.
The spherical vector in this representation is obtained by Fourier
transform on $(x_6,x_8,x_9)$,
\begin{equation}
\tilde f_{E6}=\int \frac{dx_6~dx_8~dx_9}{y^{3/2}} f_{E_6}
e^{2 i (p_6x_6+p_8x_8+p_9x_9)/y}
\end{equation}
Remarkably, the integral can still
be computed by the same method as in Section~4.1, and yields the
simple result
\begin{equation}
\label{e565}
\tilde f_{E_6}=\frac{1}{\sqrt{y J_4}} K_1\left( \frac{1}{y} \sqrt{J_4} \right)
\end{equation}
where $J_4$ is the polynomial of order 4,
\begin{equation}
J_4=y^4-\frac{y^2}{2} \Tr(X^2)+\frac18 \left((\Tr X^2)^2-2\Tr X^4\right)
\end{equation}
manifestly invariant under the maximal compact subgroup $SO(5)\subset Sl(5)$.
Note that $x_0$ is now unified with the other $x_i$ coordinates, and
that the phase has disappeared.

\subsection{$E_7$}
The same strategy presented for $E_6$ above yields the $E_7$ and $E_8$ 
spherical vectors. In the case of $E_7$, the minimal representation
has dimension 17, and is realized on a space of functions of
$(y,x_0,\dots,x_{15})$. The linearly realized subgroup
is $H_0=Sl(6,\Real)$, with maximal compact subgroup $K_0=SO(6,\Real)$.
The coordinates $(x_1,\dots,x_{15})$ transform in the adjoint
representation of $H_0$. Using the explicit expression for
the roots given in the Appendix, we can fit them into an antisymmetric matrix
\begin{equation}
\label{e57}
Z=\begin{pmatrix}
0 & -x_1 & x_2 & -x_4 & -x_6   & x_9 \\
  &  0   & x_3 & -x_5 & -x_8   & x_{12} \\
  &      & 0   &  x_7 & x_{11} & -x_{15} \\
  &      &     &   0  &-x_{14} & x_{13} \\
  & a/s     &     &      & 0      & x_{10} \\
  &      &     &      &        & 0
\end{pmatrix}
\end{equation}
The independent invariants of $Z$ under the adjoint action of $SO(6,\Real)$
are the three Casimir operators of $SO(6)\sim Sl(4)$, i.e.
\begin{equation}
\label{e58}
I_2=-\frac12\Tr(Z^2), \quad I_3=-\Pf Z,\quad I_4=\frac12\Tr(Z^4)
\end{equation}
As for $E_6$, $I_3$ is in fact invariant under the full $H_0$,
and is the cubic form that enters the expression of the Weyl generator
$A$ in~\eqref{eA}. Using the action of $K_0$, we can skew-diagonalize $Z$,
and set all coordinates but $x_1,x_7,x_{10}$ to zero. The invariants
then reduce to the simple symmetric combinations
\begin{equation}
\label{e59}
I_2=x_1^2+x_{7}^2+x_{10}^2\ ,\quad
I_3=-x_1 x_{7} x_{10}\ ,\quad
I_4=x_1^4+x_{7}^4+x_{10}^4\ .
\end{equation}
Looking at the action of $E_{\beta_{1,5,8}}$, we again find that the
spherical vector must take the form 
$f=h(S_1)e^{-i \frac{x_0 I_3}{y(y^2+x_0^2)}}/(y^2+x_0^2)^{3/2}$,
with $S_1$ the usual form 
in~\eqref{e56}. As in the $E_6$ case, it 
can be written more compactly as
\begin{equation}
\label{e592}
S_1 = \frac{\sqrt{\det( Z + |z| \mathbb{I}_6)}}{|z|^2}\, ,
\end{equation}
where again $z=y+i x_0$.

The equation $(E_{\alpha_2}+E_{-\alpha_2})f=0$ now requires
$h''+\frac{3}{S_1} h'-h=0$, hence
$h=K_1(S_1)/S_1$. The $E_7$ spherical vector is therefore given by
\begin{equation}
\label{eFE7}
f_{E_7}=\frac{K_1(S_1)}{(y^2+x_0^2)^{3/2}S_1}
e^{-i\frac{x_0 I_3}{y(x_0^2 + y^2)}}
\end{equation}
with $S_1$ as in~\eqref{e562}. In the limit $|z|\to 0$, this reduces to
\begin{equation}
\label{eFE7l}
f_{E_7}\sim \frac{e^{-\frac{|\Pf Z|}{y z}}}{|\Pf Z|^{3/2}}
\end{equation}
or its complex conjugate, depending on the sign of $\Pf Z$.

As in the $E_6$ case, we can find the spherical vector for other polarizations
as well. A particularly interesting one is obtained by Fourier transform
on the last column of the matrix $Z$ in~\eqref{e57}, which, as
examination of the root lattice shows, yields a representation with
an $SO(5,5)$ group acting linearly. The 16 coordinates now transform
as a spinor of $SO(5,5)$, or as $1+10+5$ in terms of its $Sl(5)$ subgroup,
\begin{equation}
x_0\ ,\quad
X =\begin{pmatrix}
0 & -x_1 & x_2 & -x_4 & -x_6   \\
  &  0   & x_3 & -x_5 & -x_8   \\
  &      & 0   &  x_7 & x_{11} \\
  & a/s  &     &   0  &-x_{14} \\
  &      &     &      & 0    
\end{pmatrix}\ ,\quad
Y=\begin{pmatrix}
p_9 \\ p_{12} \\ -p_{15} \\ p_{13} \\ p_{10}
\end{pmatrix}\, .
\end{equation}
Again, the Fourier transform of the spherical vector~\eqref{eFE7} can be
computed using the same method as in Section~4.1, and yields a simple
form
\begin{equation}
\label{e595}
\tilde f_{E_7}= \frac{y^{3/2}}{J_4^{5/4}} 
K_{3/2} \left( \frac{1}{y} \sqrt{J_4} \right)
\end{equation}
where $J_4$ is a $SO(5)\times SO(5)$ invariant polynomial of degree 4,
\begin{eqnarray}
J_4&=&y^4+y^2 \left( x_0^2+ Y^t Y- \frac{1}{2}\Tr X^2 \right)\\
&&+\left( x_0^2~Y^t Y
-\frac{1}{4}(\Tr X^4- \frac12(\Tr X^2)^2) -2\, x_0 X\wedge X\wedge Y 
\right) \nn
\end{eqnarray}
and the last term denotes the contraction with the five-dimensional 
Levi-Civita tensor.

\subsection{$E_8$}
Finally, in the $E_8$ case, the minimal representation has dimension $29$,
and is realized on a space of functions of $(y,x_0,\dots,x_{27})$. $E_6$
is linearly realized, and acts on $(x_1,\dots,x_{27})$ in the $27$
representation. Its maximal compact subgroup is $USp(8)$, under
which the $(x_1,\dots,x_{27})$ transform as an antitraceless antisymmetric
representation. It is somewhat awkward to fit the 27 coordinates into
an such a matrix, nevertheless we
can easily find
their transformation under the $Sl(3)\times Sl(3)\times Sl(3)$ subgroup
of $H_0=E_6$.  We have the branching rule
\begin{eqnarray}
E_6 &\supset & Sl(3)\times Sl(3)\times Sl(3) \nn\\
27 & = & (3,3,1)\oplus(3,1,3)\oplus (1,3,3)
\label{e60}
\end{eqnarray}
so that the $x_i$ can be assigned to three $3\times 3$ matrices
\begin{equation}
\label{e61}
U_{31}=-\begin{pmatrix}
x_{10}&x_{11}&x_{13}\\
x_{12}&x_{14}&x_{16}\\
x_{15}&x_{17}&x_{20}
\end{pmatrix}\ ,\quad
V_{12}=\begin{pmatrix}
x_{7}&x_{9}&x_{18}\\
-x_{6}&-x_{8}&x_{21}\\
x_{4}&x_{5}&x_{24}
\end{pmatrix}\ ,\quad
W_{23}=\begin{pmatrix}
x_{27}&-x_{25}&x_{22}\\
x_{26}&-x_{23}&x_{19}\\
x_{3}&x_{2}&x_{1}
\end{pmatrix}\ ,\quad
\end{equation}
acted upon from the left and from the right by the $Sl(3)$ factors denoted
in subscript. The maximal subgroup $K_0=USp(8)$ of $H_0=E_6$ branches
itself into $SO(3)\times SO(3)\times SO(3)$,  where the three $SU(2)$ 
are generated by $(K_{\alpha_1},K_{\alpha_3},K_{\alpha_8})$,
$(-K_{\alpha_2},K_{\alpha_{50}},K_{\alpha_{53}})$ and 
$(K_{\alpha_5},K_{\alpha_6},K_{\alpha_{12}})$ where $K_\alpha\equiv 
E_\alpha+E_{-\alpha}$,
respectively. The invariants under $K_0$ can be constructed out of
the $SO(3)\times SO(3)\times SO(3)$ invariants by requiring invariance
under the extra $K_{\alpha_4}$ compact generator,
and read
\begin{eqnarray}
\label{e62}
I_2&=& \Tr(U^t U)+\Tr(V^t V)+\Tr(W^t W)\ ,\\
I_3&=& \Tr(UVW)-(\det(U)+\det(V)+\det(W))\ ,\\
I_4&=& \Tr(U U^t U U^t)+ \Tr(V V^t V V^t)+ \Tr(W W^t W W^t) \nonumber\\
&&-2 (\Tr(U V V^t U^t)+ \Tr(V W W^t V^t)+ \Tr(W U U^t W^t) )\\
&&+2 (\Tr (U^t U)\Tr (V^t V)+\Tr (V^t V)\Tr (W^t W)+\Tr (W^t W)\Tr (U^t U))
\nonumber\\
&&+4 (\det(W) \Tr (U V W^{-t})+\det(U) \Tr (V W U^{-t})
+\det(V) \Tr (W U V^{-t}))\nonumber
\end{eqnarray}
Equivalently, we can make the $Sl(6)\times Sl(2)$ subgroup of $H_0$
manifest, by arranging the $(15,1)+(6,2)$ $x_i$'s into an antisymmetric
$6\times 6$ matrix and a doublet of 6-vectors,
\begin{equation}
\label{e622}
Z=-\begin{pmatrix}
0&x_{5}&x_{8}&x_{10}  &x_{12} &x_{15}\\
 &0    &x_9  & x_{11} &x_{14} &x_{17}\\
 &     &0    & x_{13} & x_{16} & x_{20}\\
 &     &     & 0      & x_{19} & x_{23}\\
 & a/s    &     &        & 0      & x_{26}\\
 &     &     &        &       & 0
\end{pmatrix}\ ,\quad
Y_1=\begin{pmatrix}
x_7\\-x_6\\x_4\\x_3\\x_2\\x_1\end{pmatrix}\ ,\quad
Y_2=\begin{pmatrix}
x_{18}\\x_{21}\\x_{24}\\-x_{27}\\x_{25}\\-x_{22}\end{pmatrix}\, .
\end{equation}
In this notation, the 
$K_0$-invariants can be rewritten more concisely as
\begin{eqnarray}
\label{e623}
I_2&=& -\Tr(Z^2)/2+\Tr(Y_i Y_i^t)\ ,\\
I_3&=& \Pf(Z)+ \Tr(Y_1^t Z Y_2)\ ,\\
I_4&=& \frac12 \Tr(Z^4)+ \Tr ((Y_i Y_i^t)^2)+
2\Tr Y_i^t Z^2 Y_i\\
&&- (\Tr Z^2)(\Tr Y_i Y_i^t)
+\frac12 \epsilon^{ijklmn} Z_{ij}Z_{kl}Y^1_{m} Y^2_n \, .\nonumber
\end{eqnarray}
Using an $USp(8)$ rotation, we can set all $x_i$'s to zero except e.g.,
$x_1,x_{20},x_{24}$. The invariants above then reduce to
\begin{equation}
\label{e63}
I_2=x_1^2+x_{20}^2+x_{24}^2\ ,\quad
I_3=-x_1 x_{20} x_{24}\ ,\quad
I_4=x_1^4+x_{20}^4+x_{24}^4\ .\quad
\end{equation}
The $\beta_{1,20,24}$ equations require the ansatz
\begin{equation}
\label{e64}f=(y^2+x_0^2)^{-5/2}
h(S_1) e^{-i\frac{x_0 I_3}{y(x_0^2 + y^2)}}
\end{equation}
while the $\alpha_7$ equation gives $h''+\frac{5}{S_1} h'-h=0$ and  hence
$h=K_2(S_1)/S_1^2$. The $E_8$ spherical vector is therefore
\begin{equation}
\label{efE8}
f_{E_8}=\frac{K_2(S_1)}{(y^2+x_0^2)^{5/2}S_1^2}
e^{-i\frac{x_0 I_3}{y(x_0^2 + y^2)}}
\end{equation}
with $S_1$ as in~\eqref{e56}. Again, the real part of the action $S_1$
can be more compactly written as
\begin{eqnarray}
S_1&=&\frac{1}{|z|^2}\Big[
\det(Z+|z|\mathbb{I}_6)
+|z|^4 \Tr(Y_i Y_i^t)  \\
&&+|z|^2 \left( 2\det_{\alpha\beta}( Y^i_\alpha Y^j_\beta)
+ \Tr(Y_i Y_i^t) \Tr Z^2 -\frac12 Z\wedge Z\wedge Y_1\wedge Y_2\right)\nn\\
&&+\Tr(Y_1^t Z Y_2)\left(2\Pf(Z)+ \Tr(Y_1^t Z Y_2)\right) \Big]\nn
\end{eqnarray}
As for $E_6$ and $E_7$, another interesting representation can be obtained
by Fourier transforming on the 13 coordinates $(x_0,Y_1,Y_2)$ (or, 
equivalently, under the 15 coordinates in $X$)\footnote{This representation
has also been constructed independently in~\cite{Gunaydin:2001bt}.}. 
In this polarization, 
the linearly realized symmetry group is enlarged to $Sl(8)$,
and the 28 coordinates $x_0,\dots, x_{27}$ transform as an 
antisymmetric matrix
\begin{equation}
X=-\begin{pmatrix}
0&x_{5}&x_{8}&x_{10}  &x_{12} &x_{15} &p_7 &  p_{18}\\
 &0    &x_9  & x_{11} &x_{14} &x_{17} &-p_6 & p_{21}\\
 &     &0    & x_{13} & x_{16} & x_{20} & p_4 & p_{24}\\
 &     &     & 0      & x_{19} & x_{23} &-p_3 & -p_{27}\\
 &     &     &        & 0      & x_{26} & p_2 & p_{25}\\
 &     &     &        &       & 0 & p_1 & -p_{22} \\
 & a/s& & & & &0 &p_0\\
 & & & & & & & 0
\end{pmatrix}
\end{equation}
It would be interesting to find the spherical vector in 
this representation.

\vskip 2mm
\noindent{\bf Summary}. 
The general form of the spherical invariant for $E_{6,7,8}$
in the standard minimal representation is 
\begin{eqnarray}
\label{fEn}f_{E_n}&=&\frac{{\cal K}_{s/2}(S_1)}{(y^2+x_0^2)^{(s+1)/2}}
e^{-i\frac{x_0 I_3}{y(x_0^2 + y^2)}}\nn\\
&=&\frac{1}{R^{s+1}}\,
{\cal
K}_{s/2}\Big(||(X,\nabla_{\!\!X}\Big(\frac{I_3}{R}\Big))||\Big)\, 
\exp\left(-i\, \frac{x_0I_3}{yR^2}\right)\, ,
\end{eqnarray}
where ${\cal K}_t$ is expressed in terms of the standard Bessel
function as ${\cal K}_t(x)\equiv x^{-t}K_t(x)$ and
the ``classical action'' $S_1$ is given in terms of the quadratic,
cubic and quartic invariants $I_{2,3,4}$ by
\begin{equation}
\label{e66}S_1=\frac{\sqrt{
(y^2+x_0^2)^3 + (y^2+x_0^2)^2 I_2 + (y^2+x_0^2)(I_2^2 - I_4)/2 + I_3^2}}
{y^2+x_0^2}\ =||(X,\nabla_{\!\!X}\Big(\frac{I_3}{R}\Big))||
\, ,
\end{equation}
where $X=(y,x_0,x_i)$, $R=||(y,x_0)||$ and 
$s=1,2,4$ for $n=6,7,8$ respectively, 
The parameter $s$ can be 
identified with the dimension of the field $\Real, \mathbb{C}, \mathbb{H}$
entering in the alternate construction of the minimal representations
of $E_{6,7,8}$ through Jordan algebras in~\cite{Kostant}. We have also
found alternate representations for $E_6$ and $E_7$, where $Sl(5)$
and $SO(5,5)$ act linearly, respectively; the spherical vectors in
this representation can be found in~\eqref{e565},\eqref{e595}.

\subsection{Complex spherical vectors}

For completeness, we discuss the case of a complex group,
for which our methods also allow us to derive the spherical vector.
The complex group $G(\Comp)$ can be obtained by complexifying
its split real form $G(\Real)$, {\it i.e.,} adjoining
to the real generators $E_i$ of $G(\Real)$ a set of ``imaginary'' generators
$E_i'$ such that
\begin{equation}
[T_i,T_j]=c_{ijk} T_k\ ,\quad
[T_i,T_j']=c_{ijk} T_k'\ ,\quad
[T_i',T_j']=-c_{ijk} T_k\ .
\end{equation}
Equivalently, one can introduce the holomorphic and anti-holomorphic
generators $\Tb_i=T_i+i T_i',\  \ol\Tb_i=T_i-i T_i'$, satisfying
\begin{equation}
[\Tb_i,\Tb_j]=c_{ijk} \Tb_k\ ,\quad
[\Tb_i,\ol \Tb_j]=0\ ,\quad
[\ol\Tb_i,\ol\Tb_j]=c_{ijk} \ol\Tb_k\ .\quad\end{equation}
We stress that the holomorphic generators $\Tb_i$ are identical in form to the
original $T_i$ except that the variables are now complex, while the
generators $\ol \Tb_i$ are obtained by replacing all variables by
their complex conjugates.
Dividing the generators into Cartan ones $\Hb_\a$ and those associated with
simple roots $\Eb_{\pm\a}$, the maximal compact subgroup
$K\subset G(\Comp)$ is generated by
\begin{equation}
\Eb_{\alpha}\pm\ol\Eb_{-\alpha}\ ,\qquad
\ol\Eb_{\alpha}\pm\Eb_{-\alpha}\ \quad\mbox{and}\quad
\Hb_{\a}-\ol\Hb_{\a}\, .
\end{equation}
[The choice of sign again depends on the conventions for positive and
negative roots.] $K$ is simply the real compact group of the same
type as $G$. The simplest example is $Sl(2,\Comp)$ with maximally
compact subgroup $SU(2)$. The spherical vectors in the complexified
metaplectic and Eisenstein representations 
(see~\eqref{e5},~\eqref{e5.5},~\eqref{e5.61}
and~\eqref{frog},~\eqref{elephant},~\eqref{castle} 
respectively) are
\begin{equation}
f_{meta}=e^{-x\bar x}\,, \qquad
f_{Eis}=|x|^{1-\nu}K_{\nu-1}(2\, |x|)\, .
\end{equation}

For the groups $D_4$, $E_6$, $E_7$ and $E_8$, the form of the complex spherical
vector in the standard minimal representation is uniform. Again,
the requirement that $f$ be annihilated by the linearly realized compact
subgroup allows us to reduce the problem to one in five complex variables
$(y,x_0,x_1,x_2,x_3)$ (for $E_{6,7,8}$ we have renamed the remaining $x_i$'s
for simplicity and will relabel corresponding roots accordingly).
The compact Cartan generators $\Hb_\a-\ol\Hb_\a$ imply that the all dependence
is through the complex modulus or the ratio $x_1x_2x_3/(yx_0)$, 
namely $f=f(|y|,|x_0|,|x_1|,|x_2|,|x_3|,x_1x_2x_3/(yx_0))$.
The compact generators of the root attached to the affine one
\begin{equation}
\Eb_{\beta_0}+\ol\Eb_{-\beta_0}=y\d_0-\bar x_0\bar\d+\frac{i\ol I_3}
{\bar y^2}\, ,\qquad
\ol\Eb_{\beta_0}-\Eb_{-\beta_0}=\bar y\bar \d_0-x_0\d+\frac{i I_3}{y^2}
\end{equation}
[where $\ol I_3\equiv \bar x_1\bar x_2\bar x_3+\cdots$] imply the phase 
factor
\begin{equation}
f=\exp(-iS_2)\, g(||(y,x_0)||,|x_1|,|x_2|,|x_3|)\,, 
\end{equation}
with
\begin{equation}
S_2\equiv\frac{1}{|y|^2+|x_0|^2}\Big(\frac{\bar x_0I_3}{y}+
\frac{x_0\ol I_3}{\bar y}\Big)\, .
\end{equation} 
We can drop the $x_0$ dependence of the function $g$ and
reinstate it at the end of the calculation, so we look at the root
$\beta_1$ at $x_0=0$ (and drop non-diagonal terms in the~$x_i$'s),
\begin{equation}
\Eb_{\beta_1}-\ol \Eb_{-\beta_1}=
y\d_1-\bar x_1\bar \d-\frac{\bar x_1}{\bar y}\,
\Big(s+1+\bar x_2\bar \d_2+\bar x_3\bar\d_3\Big)
\end{equation}
where the constant $s=0$ for $D_4$ and $s=1,2,4$ for $E_{6,7,8}$.
This implies (at $x_0=0)$ that 
\begin{equation}
g=|y|^{-2(s+1)}\,h\left(2\,\frac{\sqrt{(|y|^2+|x_1|^2)(|y|^2+|x_2|^2)
(|y|^2+|x_3|^2)}}{|y|^2}\right)\, .
\end{equation}
Finally, 
we examine the root $\alpha_1$ at $x_0=0$ with diagonal $x_i$'s
\begin{equation}
\Eb_{\a_1}+\ol\Eb_{-\a_1}=-\frac{ix_2x_3}{y}+\bar x_1\bar \d_0
+i\bar y\Big(\bar\d_2\bar\d_3+ \mbox{[$s$ off-diagonal double
derivatives]}\Big)\, .
\label{giraffe}
\end{equation}
Note that we may not neglect $x_0$ or off-diagonal $x_i$ derivatives
even though these variables are set to zero at the end of the calculation.
In turn the function $h$ satisfies the ordinary, Bessel-type, differential
equation $xh''+(2s+1)h'+xh=0$ (to verify that the $s$ off-diagonal
double derivative terms in~\eqref{giraffe} produce the coefficient $2s$
requires knowledge of the invariants of
the linearly realized compact subgroup described below).

It is now a matter of reinstating the dependence on the remaining
variables; orchestrating the above results we find the complex
spherical vector
\begin{equation}
f=\frac{{\cal K}_s(S_1)}{(|y|^2+|x_0|^2)^{s+1}}\, 
\exp\Big(\frac{-i\,}{|y|^2+|x_0|^2}\Big[\frac{\bar x_0I_3}{y}+
\frac{x_0\ol I_3}{\bar y}\Big]\Big)\, ,
\label{complex}
\end{equation}
where $s=0,1,2,4$ for $D_4$, $E_{6,7,8}$, respectively,
and ${\cal K}_t(x)\equiv x^{-t}K_t(x)$. The action is
\begin{equation}
S_1=2\,\frac{\sqrt{(|y|^2+|x_0|^2)^3+(|y|^2+|x_0|^2)^2I_2
+(|y|^2+|x_0|^2)(I_2^2-I_4)/2+|I_3|^2}}{|y|^2+|x_0|^2}\, ,
\end{equation}
and $I_2$,  $I_3$, $\ol I_3$ and $I_4$ are the invariants of the linearly
realized subgroup. 
For $E_6$ and $E_7$ they are subsumed by the elegant formulae
\begin{equation}
S_1^{E_6} = \frac{ \sqrt{\det( Z Z^\dagger + (|y|^2+|x_0|^2) \mathbb{I}_3) } }
{|y|^2+|x_0|^2}\, ,\qquad
S_1^{E^7} = \frac{ (\det( Z Z^\dagger + (|y|^2+|x_0|^2) \mathbb{I}_6) )^{1/4}}
{|y|^2+|x_0|^2}\, .
\end{equation}
The matrices $Z$ are the same as~\eqref{e52} and~\eqref{e57} for
complex variables; the cubic invariant $I_3$ is holomorphic and
takes the same expression as in the real case. The quadratic and
quartic invariants $I_2$ and $I_4$ have the same form as
in the real case with hermitian conjugation replacing transposition.
Finally, we note that a rewriting in terms of the norm 
$||(x_1,x_2,\ldots)||\equiv\sqrt{|x_1|^2+|x_2|^2+\cdots}$ and $R\equiv
||(y,x_0)||$ also 
holds
\begin{equation}
f=\frac{1}{R^{2(s+1)}}\,{\cal
K}_s(2\,||(X,R\,\nabla_{\!\!X}\Big(\frac{I_3}{R^2}\Big)||)
\, 
\exp\Big(\frac{-i\,}{R^2}\Big[\frac{\bar x_0I_3}{y}+
\frac{x_0\ol I_3}{\bar y}\Big]\Big)\, .
\end{equation}
Similar generalizations of the spherical vectors for 
complexifications of the
groups $A_n$ and $D_n$ may also be obtained using exactly the
methods presented above, the former is trivial, while for the $D_n$ case 
we find
\begin{equation}
\label{mould}
f_{D_n}=\frac{1}{R^2}\,\Big(\frac{||(y,x_0,x_1)||}{R}\Big)^{n-4}\,
{\cal K}_{n-4}\left(2\,||(X,R\,\nabla_{\!\!X}\Big(\frac{I_3}{R^2}\Big)||
\right)\,
\exp\Big(\frac{-i\,}{R^2}\Big[\frac{\bar x_0I_3}{y}+
\frac{x_0\ol I_3}{\bar y}\Big]\Big)\, .
\end{equation}

\section{Discussion}
The main object of this paper was the derivation of the spherical vector
for the minimal representation of real simply laced groups in the split real
form. Our results are displayed in~\eqref{fdn} and~\eqref{fEn} above. This 
spherical vector
is an essential component in the construction of automorphic theta series
for exceptional groups. It has been proposed that such objects would
provide the partition function for the winding modes of the 
quantum eleven-dimensional supermembrane in
M-theory. Although the physical interpretation of the results obtained herein
will be discussed elsewhere,  we close
with some comments, both mathematical and physical:

(i) We have obtained the spherical vector over the real field. As we
have explained in Section~2, the spherical vectors over the $p$-adic
fields $\Q_p$ are also important, since their product
gives the summation measure when constructing a theta series. 
Unfortunately in the $p$-adic case, we cannot rely on partial
differential equations anymore. One may however require both 
the spherical vector and its $A$ and $S$ transforms to have 
support on the $p$-adic integers, which together with
invariance under the linearly realized
maximal compact group $K_0(\Zint_p)$ should fix it uniquely.
Nonetheless the presentation of the real spherical vectors in terms of norms
does suggest a natural generalization to the $p$-adic case.

(ii) We have left the space of functions of $(y,x_i)$ unspecified. In
order for the generators $S$ and $A$ to be well-defined, it should consist
of functions in the Schwartz space, as well as their images. The main issue
is the regularity at the origin. A proper understanding of this issue would 
provide the degenerate contributions to the theta series which we
have overlooked in this paper.

(iii) We have only discussed the minimal representation for 
simply-laced groups. For non-simply laced cases, some differences
arise: for $G_2$ and $F_4$, the minimal representation does not 
contain any singlet under reduction to the maximal compact subgroup
$K\subset G$, therefore there is no spherical vector, but a multiplet
of such vectors transforming into each other~\cite{Vogan1}. 
It would be interesting
to find the wave function for the lowest such multiplet.
For $B_{n\geq 3}$, there
simply does not exist a representation of the same dimension as that of the
smallest nilpotent subalgebra~\cite{Vogan2}. 
For the symplectic case $C_n$, the
spherical vector is not annihilated by the compact generators, but has
a non-vanishing eigenvalue.

(iv) From a physical viewpoint, the simplicity of our results is
very encouraging. Specifically, in the $E_6$ case, we have found
a representation on $2+9$ variables, such that 9 of them transform
as a bifundamental under $Sl(3)\times Sl(3)$ in $E_6$: these
variables should be identified with the winding numbers of a membrane
on $T^3$, with the two $Sl(3)$ factors being the worldvolume and target
space modular groups, respectively. The target space $Sl(2)$ U-duality
group, corresponding to the modular transformations of the modulus
$\tau=C_{123}+i V_3$, would then be realized through Fourier transform
({\it i.e.}, Poisson resummation). It would be very interesting to understand
the physical meaning of the two extra quantum numbers $(y,x_0)$ that
we found necessary for realizing the symmetry. It would also
be important to understand if the Born-Infeld-like 
action~\eqref{e563} for the zero-modes
of the membrane can be generalized to include fluctuations, and 
yield a new description of the membrane where U-duality is 
non-linearly realized as a dynamical symmetry.
For $E_7$, we have constructed a representation in terms of 2+15 variables,
with an action~\eqref{e592} suggestive of a Born-Infeld-like action
on a six dimensional worldvolume, whose interpretation is unclear.
For $E_8$, we have a representation on 
$2+27$ variables, where $27$ of them transform linearly under $E_6$.
This is the appropriate number of charges for M-theory on $T^6$,
so the spherical vector~\eqref{efE8} should correspond to the membrane
(or, perhaps more appropriately, the five-brane)
partition function on $T^6$ in the Schwinger representation, where
the sum over BPS states is apparent. The modular group $Sl(3)$ should
then be realized by Poisson resummation. It would be very interesting
to find a representation where $Sl(3)$ acts linearly, and identify
it with the membrane action. The intertwining operator between the
two representations would then realize membrane/five-brane duality.
Finally, the representations we have displayed can be used to
construct quantum mechanical systems with spectrum-generating 
exceptional symmetries, much in the spirit of conformal 
quantum mechanics~\cite{fubini}. It would be interesting to see
if they can play a role in string theory.

\vskip 1cm

\noindent {\bf Acknowledgments.} 
The authors are grateful to K. Koepsell, H. Nicolai, N.Obers,
J. Plefka,  E. Rabinovici and N. Wyllard for useful discussions,
to D. Vogan for correspondence on the non-simply laced case, 
and especially to A. Polishchuk
for discussions about the $p$-adic case. B.P gratefully 
acknowledges the hospitality of ITP, Santa Barbara during
part of this project. B.P. and A.W. are grateful to the 
Max-Planck-Institute f\"ur Gravitationsphysik Albert-Einstein-Institut
for hospitality during the course of this project.
Many thanks are also extended to the authors of
the Form~\cite{Form}, 
Lie and Mathematica symbolic computing packages, which
have been instrumental in arriving at the results presented here.
The work of B.P. is supported by the David and Lucile Packard foundation and 
that of
A.W. by NSF grant PHY99-73935.
\appendix

\section{Group theory data}
In this appendix, we supply the list of positive roots 
for all simply-laced groups,
graded by their charge under the affine Cartan generator $H_{\omega}$.
The grade-one subspace is presented in the standard polarization,
as explained in the text below equation~\eqref{e20}, and the action of
the Weyl reflection $A$ w.r.t. to $\beta_0$ is indicated. 
The explicit expressions for the cubic invariant $I_3$, 
the Cartan generators and the (positive and
negative) simple roots in the minimal representation are listed.
The expressions for the grade 1 and 2 positive roots are given in~\eqref{e21},
and repeated here for convenience,
\begin{equation}
E_{\beta_i}=y \partial_i\ ,\quad E_{\gamma_i}=i x_i\ ,\quad
E_{\omega}=i y\ , \qquad i=0,\dots,d-1\ .
\end{equation}
The Cartan generators for the simple root $\beta_0$ and the affine root
$\omega$ take also universal forms,
\begin{eqnarray}
H_{\beta_0}&=&- y \partial + x_{0} \partial_{0}\\
H_{\omega}&=&-\nu -2y\pa -\sum_{i=0}^{d-1} x_i\partial_{i}
\end{eqnarray}
up to the ``normal ordering constant'' $\nu=(n-1),6,9,15$ for 
$D_n,E_6,E_7,E_8$, respectively.
More compact expressions can be obtained by making manifest
the covariance under the linearly realized group $H_0$.

\subsection{$A_n$}
Dynkin diagram:

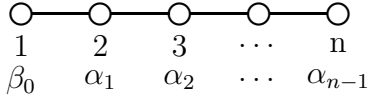
\begin{picture}(190,30)
\thicklines
\multiput(0,0)(30,0){5}{\circle{8}}
\put(0,-12){\makebox(0,0){1}}\put(0,-25){\makebox(0,0){$\beta_0$}}
\put(30,-12){\makebox(0,0){2}}\put(30,-25){\makebox(0,0){$\alpha_1$}}
\put(60,-12){\makebox(0,0){3}}\put(60,-25){\makebox(0,0){$\alpha_2$}}
\put(90,-12){\makebox(0,0){$\dots$}}\put(90,-25){\makebox(0,0){$\dots$}}
\put(120,-12){\makebox(0,0){n}}\put(120,-25){\makebox(0,0){$\alpha_{n-1}$}}
\multiput(4,0)(30,0){4}{\line(1,0){22}}
\end{picture}

\vskip 1cm
\noindent Positive roots:
$$\begin{array}{rccccccccc}
\alpha_1&= &   (0,&1,&0,&\dots ,& 0,&0&)     &=A(\beta_1)\\
\alpha_2&= &   (0,&0,&1,&\dots ,& 0,&0&)     &\\
\vdots  &  &   (0,&0,&0,&\ddots,& 0,&0&)     &\\
\alpha_{n-2}&= &   (0,&0,&0,&\dots ,& 1,&0&) &\\
\alpha_{n-1}&= &  (0,&1,&1,&0,&\dots,&0&)     &=A(\beta_2)\\
\alpha_{n}&=   &  (0,&0,&1,&1,&0,&0&)     &\\
\vdots  &  &   (0,&0,&0,&\ddots,&\ddots,&0&)  &\\
\alpha_{2n-5}&=   &  (0,&0,&\dots,&1,&1,&0&)     &\\
\vdots    & & \\
\alpha_{(n-1)(n-2)/2}&=   &  (0,&1,&\dots,&1,&1,&0&)     &=A(\beta_{n-2})\\
\end{array}$$
$$\begin{array}{ccccccc@{\hspace{7mm}}ccccccc}
\beta_0&= &  (1,&0,&0,&0,&0)  &   \gamma_0&=&(0,&1,&\dots,&1,&1)\\
\beta_1&= &  (1,&1,&0,&0,&0)  &   \gamma_1&=&(0,&0,&1,&\dots,&1)\\
\vdots   &&  (\vdots,&\vdots ,&\ddots,&0,&0)&\vdots&&(0,&0,&\dots,&\ddots,&1)\\
\beta_{n-2}&= &   (1,&1,&\dots,&1,&0)  &   \gamma_{n-2}&=&(0,&0,&\dots,&0,&1)
\end{array}$$
$$\begin{array}{ccc}
\omega = &   (1,1,\dots,1,1)  & =A(\gamma_0)
\end{array}$$
\noindent Cartan generators ($\nu=(n+1)/2$ in the standard minimal rep):
\begin{eqnarray*}
H_{\beta_0}&=&- y \partial + x_{0} \partial_{0}\\
H_{\alpha_1}&=&- x_{0} \partial_{0}+ x_{1} \partial_{1}\\
H_{\alpha_2}&=&- x_{1} \partial_{1}+ x_{2} \partial_{2}\\
\vdots&&\\
H_{\alpha_{n-2}}&=&- x_{n-3} \partial_{n-3}+ x_{n-2} \partial_{n-2}\\
H_{\gamma_{n-2}}&=&- \nu - y\pa - x_0\pa_0 - \dots - x_{n-3}\pa_{n-3} 
- 2x_{n-2}\pa_{n-2}
\end{eqnarray*}
\noindent Simple roots:
\begin{equation*}
\begin{array}{ccc@{\hspace{7mm}}ccc}
E_{\alpha_{1}}&=&x_0\pa_1\ , & E_{-\alpha_{1}}&=&x_1\pa_0\\
E_{\alpha_{2}}&=&x_1\pa_2\ , & E_{-\alpha_{2}}&=&x_2\pa_1\\
\vdots &=&\vdots&\vdots &=&\vdots\\
E_{\alpha_{n-2}}&=&x_{n-3}\pa_{n-2}\ ,&
E_{-\alpha_{n-2}}&=&x_{n-2}\pa_{n-3}
\end{array}
\end{equation*}

\subsection{$D_5$}
Dynkin diagram:

\begin{picture}(190,50)
\thicklines
\multiput(0,0)(30,0){4}{\circle{8}}
\put(0,-12){\makebox(0,0){1}}\put(0,-25){\makebox(0,0){$\alpha_1$}}
\put(30,-12){\makebox(0,0){2}}\put(30,-25){\makebox(0,0){$\beta_0$}}
\put(60,-12){\makebox(0,0){3}}\put(60,-25){\makebox(0,0){$\alpha_2$}}
\put(90,-12){\makebox(0,0){5}}\put(90,-25){\makebox(0,0){$\alpha_4$}}
\multiput(4,0)(30,0){3}{\line(1,0){22}}
\put(60,4){\line(0,1){22}}
\put(60,30){\circle{8}}
\put(48,30){\makebox(0,0){4}}\put(75,30){\makebox(0,0){$\alpha_3$}}
\end{picture}


\vskip 1cm

\noindent Positive roots:
$$\begin{array}{ccc}
\alpha_1= &   (1,0,0,0,0) & =A(\beta_1)\\
\alpha_2= &   (0,0,1,0,0) & =A(\beta_2)\\
\alpha_3= &   (0,0,0,1,0) & =A(\alpha_3)\\
\alpha_4= &   (0,0,0,0,1) & =A(\alpha_4)\\
\alpha_5= &   (0,0,1,1,0) & =A(\beta_4)\\
\alpha_6= &   (0,0,1,0,1) & =A(\beta_5)\\
\alpha_7= &   (0,0,1,1,1) & =A(\beta_3)
\end{array}$$
$$\begin{array}{cc@{\hspace{7mm}}c}
\beta_0= &   (0,1,0,0,0)  &   \gamma_0=(1,1,2,1,1)\\
\beta_1= &   (1,1,0,0,0)  &   \gamma_1=(0,1,2,1,1)\\
\beta_2= &   (0,1,1,0,0)  &   \gamma_2=(1,1,1,1,1)\\
\beta_3= &   (0,1,1,1,1)  &   \gamma_3=(1,1,1,0,0)\\
\beta_4= &   (0,1,1,1,0)  &   \gamma_4=(1,1,1,0,1)\\
\beta_5= &   (0,1,1,0,1)  &   \gamma_5=(1,1,1,1,0)
\end{array}$$
$$\begin{array}{ccc}
\omega = &   (1,2,2,1,1)  & =A(\gamma_0)
\end{array}$$
\noindent Cubic form:
$$I_3= x_{1} ( x_{2} x_{3} - x_{4} x_{5} )$$
\noindent Cartan generators:
\begin{eqnarray*}
H_{\beta_0}&=&- y \partial + x_{0} \partial_{0}\\
H_{\alpha_1}&=&- 2 - x_{0} \partial_{0} + x_{1} \partial_{1} - x_{2} \partial_{2} - x_{3} \partial_{3} - x_{4} \partial_{4} - x_{5} \partial_{5}\\
H_{\alpha_2}&=&- 1 - x_{0} \partial_{0} - x_{1} \partial_{1} + x_{2} \partial_{2} - x_{3} \partial_{3}\\
H_{\alpha_3}&=&- x_{2} \partial_{2} + x_{3} \partial_{3} + x_{4} \partial_{4} - x_{5} \partial_{5}\\
H_{\alpha_4}&=&- x_{2} \partial_{2} + x_{3} \partial_{3} - x_{4} \partial_{4} + x_{5} \partial_{5}
\end{eqnarray*}
\noindent Simple roots:
\begin{eqnarray*}
E_{\alpha_{1}}&=&- x_{0} \partial_{1} - i (x_{2} x_{3} - x_{4} x_{5})/y \\
E_{\alpha_{2}}&=&- x_{0} \partial_{2} - i x_{1} x_{3}/y \\
E_{\alpha_{3}}&=&x_{2} \partial_{4} + x_{5} \partial_{3}\\
E_{\alpha_{4}}&=&- x_{2} \partial_{5} - x_{4} \partial_{3}\\
E_{-\alpha_{1}}&=& x_{1} \partial_{0}+ i y~(\partial_{2} \partial_{3} - \partial_{4} \partial_{5}) \\
E_{-\alpha_{2}}&=&x_{2} \partial_{0}+ i y~\partial_{1} \partial_{3}\\
E_{-\alpha_{3}}&=&- x_{3} \partial_{5} - x_{4} \partial_{2}\\
E_{-\alpha_{4}}&=&x_{3} \partial_{4} + x_{5} \partial_{2}
\end{eqnarray*}

\subsection{$E_6$}
Dynkin diagram:

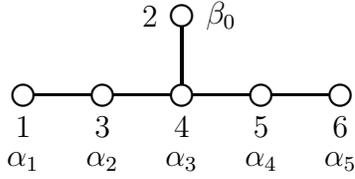
\begin{picture}(190,50)
\thicklines
\multiput(0,0)(30,0){5}{\circle{8}}
\put(0,-12){\makebox(0,0){1}}\put(0,-25){\makebox(0,0){$\alpha_1$}}
\put(30,-12){\makebox(0,0){3}}\put(30,-25){\makebox(0,0){$\alpha_2$}}
\put(60,-12){\makebox(0,0){4}}\put(60,-25){\makebox(0,0){$\alpha_3$}}
\put(90,-12){\makebox(0,0){5}}\put(90,-25){\makebox(0,0){$\alpha_4$}}
\put(120,-12){\makebox(0,0){6}}\put(120,-25){\makebox(0,0){$\alpha_5$}}
\multiput(4,0)(30,0){4}{\line(1,0){22}}
\put(60,4){\line(0,1){22}}
\put(60,30){\circle{8}}
\put(48,30){\makebox(0,0){2}}\put(75,30){\makebox(0,0){$\beta_0$}}
\end{picture}
\vskip 1cm


\noindent Positive roots:
$$\begin{array}{ccc}
\alpha_1= &      (1,0,0,0,0,0) & =A(\alpha_1)\\
\alpha_2= &      (0,0,1,0,0,0) & =A(\alpha_2)\\
\alpha_3= &      (0,0,0,1,0,0) & =A(\beta_1)\\
\alpha_4= &      (0,0,0,0,1,0) & =A(\alpha_4)\\
\alpha_5= &      (0,0,0,0,0,1) & =A(\alpha_5)\\
\alpha_6= &      (1,0,1,0,0,0) & =A(\alpha_6)\\
\alpha_7= &      (0,0,1,1,0,0) & =A(\beta_2)\\
\end{array}$$
$$\begin{array}{ccc}
\alpha_8= &      (0,0,0,1,1,0) & =A(\beta_3)\\
\alpha_9= &      (0,0,0,0,1,1) & =A(\alpha_9)\\
\alpha_{10}= &   (1,0,1,1,0,0) & =A(\beta_4)\\
\alpha_{11}= &   (0,0,1,1,1,0) & =A(\beta_5)\\
\alpha_{12}= &   (0,0,0,1,1,1) & =A(\beta_6)\\
\alpha_{13}= &   (1,0,1,1,1,0) & =A(\beta_7)\\
\alpha_{14}= &   (0,0,1,1,1,1) & =A(\beta_9)\\
\alpha_{15}= &   (1,0,1,1,1,1) & =A(\beta_8)
\end{array}$$
$$\begin{array}{cc@{\hspace{7mm}}c}
\beta_0= &   (0,1,0,0,0,0)  &   \gamma_0=(1,1,2,3,2,1)\\
\beta_1= &   (0,1,0,1,0,0)  &   \gamma_1=(1,1,2,2,2,1)\\
\beta_2= &   (0,1,1,1,0,0)  &   \gamma_2=(1,1,1,2,2,1)\\
\beta_3= &   (0,1,0,1,1,0)  &   \gamma_3=(1,1,2,2,1,1)\\
\beta_4= &   (1,1,1,1,0,0)  &   \gamma_4=(0,1,1,2,2,1)\\
\beta_5= &   (0,1,1,1,1,0)  &   \gamma_5=(1,1,1,2,1,1) \\
\beta_6= &   (0,1,0,1,1,1)  &   \gamma_6=(1,1,2,2,1,0) \\
\beta_7= &   (1,1,1,1,1,0)  &   \gamma_7=(0,1,1,2,1,1) \\
\beta_8= &   (1,1,1,1,1,1)  &   \gamma_8=(0,1,1,2,1,0) \\
\beta_9= &   (0,1,1,1,1,1)  &   \gamma_9=(1,1,1,2,1,0) 
\end{array}$$
$$\begin{array}{ccc}
\omega = &   (1,2,2,3,2,1)  & =A(\gamma_0)
\end{array}$$
\noindent Cubic form:
$$I_3=  -  x_{1} x_{5} x_{8} +  x_{1} x_{7} x_{9} +  x_{2} x_{3}  x_{8} -
x_{2} x_{6} x_{7} -  x_{3} x_{4} x_{9} +   x_{4} x_{5} x_{6} $$
\noindent Cartan generators:
\begin{eqnarray*}
H_{\beta_0}&=&- y \pa + x_{0} \pa_{0}\\
H_{\alpha_1}&=& - x_{2} \pa_{2} + x_{4} \pa_{4} - x_{5} \pa_{5} + x_{7} \pa_{7} + x_{8} \pa_{8} - x_{9} \pa_{9}\\
H_{\alpha_2}&=& - x_{1} \pa_{1} + x_{2} \pa_{2} - x_{3} \pa_{3} + x_{5} \pa_{5} - x_{6} \pa_{6} + x_{9} \pa_{9}\\
H_{\alpha_3}&=& - 2 - x_{0} \pa_{0} + x_{1} \pa_{1} - x_{5} \pa_{5} - x_{7} \pa_{7} - x_{8} \pa_{8} - x_{9} \pa_{9}\\
H_{\alpha_4}&=& - x_{1} \pa_{1} - x_{2} \pa_{2} + x_{3} \pa_{3} - x_{4} \pa_{4} + x_{5} \pa_{5} + x_{7} \pa_{7}\\
H_{\alpha_5}&=& - x_{3} \pa_{3} - x_{5} \pa_{5} + x_{6} \pa_{6} - x_{7} \pa_{7} + x_{8} \pa_{8} + x_{9} \pa_{9}\\
\end{eqnarray*}
\noindent Simple roots:
\begin{eqnarray*}
E_{\alpha_{1}}&=&- x_{2} \pa_{4} - x_{5} \pa_{7} - x_{9} \pa_{8}\\
E_{\alpha_{2}}&=&- x_{1} \pa_{2} - x_{3} \pa_{5} - x_{6} \pa_{9}\\
E_{\alpha_{3}}&=&- x_{0} \pa_{1}+ i (x_{5} x_{8}  - x_{7} x_{9})/y\\
E_{\alpha_{4}}&=&- x_{1} \pa_{3} - x_{2} \pa_{5} - x_{4} \pa_{7}\\
E_{\alpha_{5}}&=&- x_{3} \pa_{6} - x_{5} \pa_{9} - x_{7} \pa_{8}\\
E_{-\alpha_{1}}&=&x_{4} \pa_{2} + x_{7} \pa_{5} + x_{8} \pa_{9}\\
E_{-\alpha_{2}}&=&x_{2} \pa_{1} + x_{5} \pa_{3} + x_{9} \pa_{6}\\
E_{-\alpha_{3}}&=& x_{1} \pa_{0}- iy (\pa_{5} \pa_{8}  -\pa_{7} \pa_{9})\\ 
E_{-\alpha_{4}}&=&x_{3} \pa_{1} + x_{5} \pa_{2} + x_{7} \pa_{4}\\
E_{-\alpha_{5}}&=&x_{6} \pa_{3} + x_{8} \pa_{7} + x_{9} \pa_{5}
\end{eqnarray*}

\subsection{$E_7$}
Dynkin diagram:

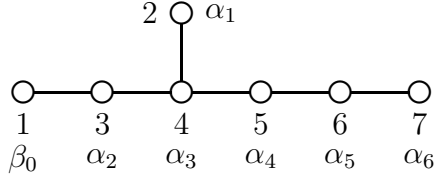
\begin{picture}(190,70)
\thicklines
\multiput(0,0)(30,0){6}{\circle{8}}
\put(0,-12){\makebox(0,0){1}}\put(0,-25){\makebox(0,0){$\beta_0$}}
\put(30,-12){\makebox(0,0){3}}\put(30,-25){\makebox(0,0){$\alpha_2$}}
\put(60,-12){\makebox(0,0){4}}\put(60,-25){\makebox(0,0){$\alpha_3$}}
\put(90,-12){\makebox(0,0){5}}\put(90,-25){\makebox(0,0){$\alpha_4$}}
\put(120,-12){\makebox(0,0){6}}\put(120,-25){\makebox(0,0){$\alpha_5$}}
\put(150,-12){\makebox(0,0){7}}\put(150,-25){\makebox(0,0){$\alpha_6$}}
\multiput(4,0)(30,0){5}{\line(1,0){22}}
\put(60,4){\line(0,1){22}}
\put(60,30){\circle{8}}
\put(48,30){\makebox(0,0){2}}\put(75,30){\makebox(0,0){$\alpha_1$}}
\end{picture}


\vskip 1cm

\noindent Positive roots:
$$\begin{array}{ccc}
\alpha_1= &        (0,1,0,0,0,0,0) & =A(\alpha_1)\\
\alpha_2= &        (0,0,1,0,0,0,0) & =A(\beta_1)\\
\alpha_3= &        (0,0,0,1,0,0,0) & =A(\alpha_3)\\
\alpha_4= &        (0,0,0,0,1,0,0) & =A(\alpha_4)\\
\alpha_5= &        (0,0,0,0,0,1,0) & =A(\alpha_5)\\
\alpha_6= &        (0,0,0,0,0,0,1) & =A(\alpha_6)\\
\alpha_7= &        (0,1,0,1,0,0,0) & =A(\alpha_7)\\
\alpha_8= &        (0,0,1,1,0,0,0) & =A(\beta_2)\\
\alpha_9= &        (0,0,0,1,1,0,0) & =A(\alpha_9)\\
\end{array}$$
$$\begin{array}{ccc}
\alpha_{10}= &     (0,0,0,0,1,1,0) & =A(\alpha_{10})\\
\alpha_{11}= &     (0,0,0,0,0,1,1) & =A(\alpha_{11})\\
\alpha_{12}= &     (0,1,1,1,0,0,0) & =A(\beta_3)\\
\alpha_{13}= &     (0,1,0,1,1,0,0) & =A(\alpha_{13})\\
\alpha_{14}= &     (0,0,1,1,1,0,0) & =A(\beta_4)\\
\alpha_{15}= &     (0,0,0,1,1,1,0) & =A(\alpha_{15})\\
\alpha_{16}= &     (0,0,0,0,1,1,1) & =A(\alpha_{16})\\
\alpha_{17}= &     (0,1,1,1,1,0,0) & =A(\beta_5)\\
\alpha_{18}= &     (0,1,0,1,1,1,0) & =A(\alpha_{18})\\
\alpha_{19}= &     (0,0,1,1,1,1,0) & =A(\beta_6)\\
\alpha_{20}= &     (0,0,0,1,1,1,1) & =A(\alpha_{20})\\
\alpha_{21}= &     (0,1,1,2,1,0,0) & =A(\beta_7)\\
\alpha_{22}= &     (0,1,1,1,1,1,0) & =A(\beta_8)\\
\alpha_{23}= &     (0,1,0,1,1,1,1) & =A(\alpha_{23})\\
\alpha_{24}= &     (0,0,1,1,1,1,1) & =A(\beta_9)\\
\alpha_{25}= &     (0,1,1,2,1,1,0) & =A(\beta_{11})\\
\alpha_{26}= &     (0,1,1,1,1,1,1) & =A(\beta_{12})\\
\alpha_{27}= &     (0,1,1,2,2,1,0) & =A(\beta_{14})\\
\alpha_{28}= &     (0,1,1,2,1,1,1) & =A(\beta_{15})\\
\alpha_{29}= &     (0,1,1,2,2,1,1) & =A(\beta_{13})\\
\alpha_{30}= &     (0,1,1,2,2,2,1) & =A(\beta_{10})\\

\end{array}$$
$$\begin{array}{cc@{\hspace{7mm}}c}
\beta_{0}= &   (1,0,0,0,0,0,0)  &   \gamma_{0}=(1,2,3,4,3,2,1)\\
\beta_{1}= &   (1,0,1,0,0,0,0)  &   \gamma_{1}=(1,2,2,4,3,2,1)\\
\beta_{2}= &   (1,0,1,1,0,0,0)  &   \gamma_{2}=(1,2,2,3,3,2,1)\\
\beta_{3}= &   (1,1,1,1,0,0,0)  &   \gamma_{3}=(1,1,2,3,3,2,1)\\
\beta_{4}= &   (1,0,1,1,1,0,0)  &   \gamma_{4}=(1,2,2,3,2,2,1)\\
\beta_{5}= &   (1,1,1,1,1,0,0)  &   \gamma_{5}=(1,1,2,3,2,2,1)\\
\beta_{6}= &   (1,0,1,1,1,1,0)  &   \gamma_{6}=(1,2,2,3,2,1,1)\\
\beta_{7}= &   (1,1,1,2,1,0,0)  &   \gamma_{7}=(1,1,2,2,2,2,1)\\
\beta_{8}= &   (1,1,1,1,1,1,0)  &   \gamma_{8}=(1,1,2,3,2,1,1)\\
\beta_{9}= &   (1,0,1,1,1,1,1)  &   \gamma_{9}=(1,2,2,3,2,1,0)\\
\beta_{10}= &   (1,1,1,2,2,2,1)  &   \gamma_{10}=(1,1,2,2,1,0,0)\\
\beta_{11}= &   (1,1,1,2,1,1,0)  &   \gamma_{11}=(1,1,2,2,2,1,1)\\     
\beta_{12}= &   (1,1,1,1,1,1,1)  &   \gamma_{12}=(1,1,2,3,2,1,0)\\
\beta_{13}= &   (1,1,1,2,2,1,1)  &   \gamma_{13}=(1,1,2,2,1,1,0)\\     
\beta_{14}= &   (1,1,1,2,2,1,0)  &   \gamma_{14}=(1,1,2,2,1,1,1)\\
\beta_{15}= &   (1,1,1,2,1,1,1)  &   \gamma_{15}=(1,1,2,2,2,1,0)\\
\end{array}$$
$$\begin{array}{ccc}
\omega = &   (2,2,3,4,3,2,1)  & =A(\gamma_0)
\end{array}$$
\noindent Cubic form:
\begin{eqnarray*}
I_3&=&-x_{1}x_{7}x_{10}+x_{1}x_{11}x_{13}-x_{1}x_{14}x_{15}+x_{2}x_{5}x_{10}-x_{2}x_{8}x_{13}+x_{2}x_{12}x_{14}-x_{3}x_{4}x_{10}+x_{3}x_{6}x_{13}\\
&&-x_{3}x_{9}x_{14}+x_{4}x_{8}x_{15}-x_{4}x_{11}x_{12}-x_{5}x_{6}x_{15}+x_{5}x_{9}x_{11}+x_{6}x_{7}x_{12}-x_{7}x_{8}x_{9}
\end{eqnarray*}
\noindent Cartan generators:
\begin{eqnarray*}
H_{\beta_0}&=&-y\pa+x_{0}\pa_{0}\\
H_{\alpha_1}&=&-x_{2}\pa_{2}+x_{3}\pa_{3}-x_{4}\pa_{4}+x_{5}\pa_{5}-x_{6}\pa_{6}+x_{8}\pa_{8}-x_{9}\pa_{9}+x_{12}\pa_{12}\\
H_{\alpha_2}&=&-3-x_{0}\pa_{0}+x_{1}\pa_{1}-x_{7}\pa_{7}-x_{10}\pa_{10}-x_{11}\pa_{11}-x_{13}\pa_{13}-x_{14}\pa_{14}-x_{15}\pa_{15}\\
H_{\alpha_3}&=&-x_{1}\pa_{1}+x_{2}\pa_{2}-x_{5}\pa_{5}+x_{7}\pa_{7}-x_{8}\pa_{8}+x_{11}\pa_{11}-x_{12}\pa_{12}+x_{15}\pa_{15}\\
H_{\alpha_4}&=&-x_{2}\pa_{2}-x_{3}\pa_{3}+x_{4}\pa_{4}+x_{5}\pa_{5}-x_{11}\pa_{11}+x_{13}\pa_{13}+x_{14}\pa_{14}-x_{15}\pa_{15}\\
H_{\alpha_5}&=&-x_{4}\pa_{4}-x_{5}\pa_{5}+x_{6}\pa_{6}-x_{7}\pa_{7}+x_{8}\pa_{8}+x_{10}\pa_{10}+x_{11}\pa_{11}-x_{13}\pa_{13}\\
H_{\alpha_6}&=&-x_{6}\pa_{6}-x_{8}\pa_{8}+x_{9}\pa_{9}-x_{11}\pa_{11}+x_{12}\pa_{12}+x_{13}\pa_{13}-x_{14}\pa_{14}+x_{15}\pa_{15}
\end{eqnarray*}
\noindent Simple roots:
\begin{eqnarray*}
E_{\alpha_{1}}&=&x_{2}\pa_{3}+x_{4}\pa_{5}+x_{6}\pa_{8}+x_{9}\pa_{12}\\
E_{\alpha_{2}}&=&-x_{0}\pa_{1}+i(x_{7}x_{10}-x_{11}x_{13}+x_{14}x_{15})/y\\
E_{\alpha_{3}}&=&x_{1}\pa_{2}+x_{5}\pa_{7}+x_{8}\pa_{11}+x_{12}\pa_{15}\\
E_{\alpha_{4}}&=&x_{2}\pa_{4}+x_{3}\pa_{5}+x_{11}\pa_{14}+x_{15}\pa_{13}\\
E_{\alpha_{5}}&=&x_{4}\pa_{6}+x_{5}\pa_{8}+x_{7}\pa_{11}+x_{13}\pa_{10}\\
E_{\alpha_{6}}&=&x_{6}\pa_{9}+x_{8}\pa_{12}+x_{11}\pa_{15}+x_{14}\pa_{13}\\
E_{-\alpha_{1}}&=&-x_{3}\pa_{2}-x_{5}\pa_{4}-x_{8}\pa_{6}-x_{12}\pa_{9}\\
E_{-\alpha_{2}}&=&x_{1}\pa_{0}-iy(\pa_{7}\pa_{10}-\pa_{11}\pa_{13}+\pa_{14}\pa_{15})\\
E_{-\alpha_{3}}&=&-x_{2}\pa_{1}-x_{7}\pa_{5}-x_{11}\pa_{8}-x_{15}\pa_{12}\\
E_{-\alpha_{4}}&=&-x_{4}\pa_{2}-x_{5}\pa_{3}-x_{13}\pa_{15}-x_{14}\pa_{11}\\
E_{-\alpha_{5}}&=&-x_{6}\pa_{4}-x_{8}\pa_{5}-x_{10}\pa_{13}-x_{11}\pa_{7}\\
E_{-\alpha_{6}}&=&-x_{9}\pa_{6}-x_{12}\pa_{8}-x_{13}\pa_{14}-x_{15}\pa_{11}
\end{eqnarray*}

\subsection{$E_8$}
Dynkin diagram:

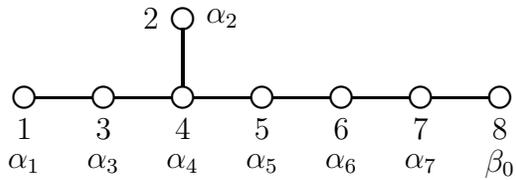
\begin{picture}(190,50)
\thicklines
\multiput(0,0)(30,0){7}{\circle{8}}
\put(0,-12){\makebox(0,0){1}}\put(0,-25){\makebox(0,0){$\alpha_1$}}
\put(30,-12){\makebox(0,0){3}}\put(30,-25){\makebox(0,0){$\alpha_3$}}
\put(60,-12){\makebox(0,0){4}}\put(60,-25){\makebox(0,0){$\alpha_4$}}
\put(90,-12){\makebox(0,0){5}}\put(90,-25){\makebox(0,0){$\alpha_5$}}
\put(120,-12){\makebox(0,0){6}}\put(120,-25){\makebox(0,0){$\alpha_6$}}
\put(150,-12){\makebox(0,0){7}}\put(150,-25){\makebox(0,0){$\alpha_7$}}
\put(180,-12){\makebox(0,0){8}}\put(180,-25){\makebox(0,0){$\beta_0$}}
\multiput(4,0)(30,0){6}{\line(1,0){22}}
\put(60,4){\line(0,1){22}}
\put(60,30){\circle{8}}
\put(48,30){\makebox(0,0){2}}\put(75,30){\makebox(0,0){$\alpha_2$}}
\end{picture}

\noindent Positive roots:
$$\begin{array}{ccc}
\alpha_{1}= &(1,0,0,0,0,0,0,0) & =A(\alpha_{1})\\
\alpha_{2}= &(0,1,0,0,0,0,0,0) & =A(\alpha_{2})\\
\alpha_{3}= &(0,0,1,0,0,0,0,0) & =A(\alpha_{3})\\
\alpha_{4}= &(0,0,0,1,0,0,0,0) & =A(\alpha_{4})\\
\alpha_{5}= &(0,0,0,0,1,0,0,0) & =A(\alpha_{5})\\
\alpha_{6}= &(0,0,0,0,0,1,0,0) & =A(\alpha_{6})\\
\alpha_{7}= &(0,0,0,0,0,0,1,0) & =A(\beta_{1})\\
\alpha_{8}= &(1,0,1,0,0,0,0,0) & =A(\alpha_{8})\\
\alpha_{9}= &(0,1,0,1,0,0,0,0) & =A(\alpha_{9})\\
\alpha_{10}= &(0,0,1,1,0,0,0,0) & =A(\alpha_{10})\\
\alpha_{11}= &(0,0,0,1,1,0,0,0) & =A(\alpha_{11})\\
\alpha_{12}= &(0,0,0,0,1,1,0,0) & =A(\alpha_{12})\\
\alpha_{13}= &(0,0,0,0,0,1,1,0) & =A(\beta_{2})\\
\alpha_{14}= &(1,0,1,1,0,0,0,0) & =A(\alpha_{14})\\
\alpha_{15}= &(0,1,1,1,0,0,0,0) & =A(\alpha_{15})\\
\alpha_{16}= &(0,1,0,1,1,0,0,0) & =A(\alpha_{16})\\
\alpha_{17}= &(0,0,1,1,1,0,0,0) & =A(\alpha_{17})\\
\alpha_{18}= &(0,0,0,1,1,1,0,0) & =A(\alpha_{18})\\
\alpha_{19}= &(0,0,0,0,1,1,1,0) & =A(\beta_{3})\\
\alpha_{20}= &(1,1,1,1,0,0,0,0) & =A(\alpha_{20})\\
\alpha_{21}= &(1,0,1,1,1,0,0,0) & =A(\alpha_{21})\\
\alpha_{22}= &(0,1,1,1,1,0,0,0) & =A(\alpha_{22})\\
\alpha_{23}= &(0,1,0,1,1,1,0,0) & =A(\alpha_{23})\\
\alpha_{24}= &(0,0,1,1,1,1,0,0) & =A(\alpha_{24})\\
\alpha_{25}= &(0,0,0,1,1,1,1,0) & =A(\beta_{4})\\
\alpha_{26}= &(1,1,1,1,1,0,0,0) & =A(\alpha_{26})\\
\alpha_{27}= &(1,0,1,1,1,1,0,0) & =A(\alpha_{27})\\
\alpha_{28}= &(0,1,1,2,1,0,0,0) & =A(\alpha_{28})\\
\alpha_{29}= &(0,1,1,1,1,1,0,0) & =A(\alpha_{29})\\ 
\alpha_{30}= &(0,1,0,1,1,1,1,0) & =A(\beta_{5})\\
\alpha_{31}= &(0,0,1,1,1,1,1,0) & =A(\beta_{6})\\
\alpha_{32}= &(1,1,1,2,1,0,0,0) & =A(\alpha_{32})\\
\alpha_{33}= &(1,1,1,1,1,1,0,0) & =A(\alpha_{33})\\
\alpha_{34}= &(1,0,1,1,1,1,1,0) & =A(\beta_{7})\\
\alpha_{35}= &(0,1,1,2,1,1,0,0) & =A(\alpha_{35})\\
\alpha_{36}= &(0,1,1,1,1,1,1,0) & =A(\beta_{8})\\
\alpha_{37}= &(1,1,2,2,1,0,0,0) & =A(\alpha_{37})\\
\alpha_{38}= &(1,1,1,2,1,1,0,0) & =A(\alpha_{38})\\
\alpha_{39}= &(1,1,1,1,1,1,1,0) & =A(\beta_{9})\\
\end{array}$$
$$\begin{array}{ccc}
\alpha_{40}= &(0,1,1,2,2,1,0,0) & =A(\alpha_{40})\\
\alpha_{41}= &(0,1,1,2,1,1,1,0) & =A(\beta_{10})\\
\alpha_{42}= &(1,1,2,2,1,1,0,0) & =A(\alpha_{42})\\
\alpha_{43}= &(1,1,1,2,2,1,0,0) & =A(\alpha_{43})\\
\alpha_{44}= &(1,1,1,2,1,1,1,0) & =A(\beta_{11})\\
\alpha_{45}= &(0,1,1,2,2,1,1,0) & =A(\beta_{12})\\
\alpha_{46}= &(1,1,2,2,2,1,0,0) & =A(\alpha_{46})\\
\alpha_{47}= &(1,1,2,2,1,1,1,0) & =A(\beta_{13})\\
\alpha_{48}= &(1,1,1,2,2,1,1,0) & =A(\beta_{14})\\
\alpha_{49}= &(0,1,1,2,2,2,1,0) & =A(\beta_{15})\\
\alpha_{50}= &(1,1,2,3,2,1,0,0) & =A(\alpha_{50})\\
\alpha_{51}= &(1,1,2,2,2,1,1,0) & =A(\beta_{16})\\
\alpha_{52}= &(1,1,1,2,2,2,1,0) & =A(\beta_{17})\\
\alpha_{53}= &(1,2,2,3,2,1,0,0) & =A(\alpha_{53})\\
\alpha_{54}= &(1,1,2,3,2,1,1,0) & =A(\beta_{19})\\
\alpha_{55}= &(1,1,2,2,2,2,1,0) & =A(\beta_{20})\\
\alpha_{56}= &(1,2,2,3,2,1,1,0) & =A(\beta_{22})\\
\alpha_{57}= &(1,1,2,3,2,2,1,0) & =A(\beta_{23})\\
\alpha_{58}= &(1,2,2,3,2,2,1,0) & =A(\beta_{25})\\
\alpha_{59}= &(1,1,2,3,3,2,1,0) & =A(\beta_{26})\\
\alpha_{60}= &(1,2,2,3,3,2,1,0) & =A(\beta_{27})\\
\alpha_{61}= &(1,2,2,4,3,2,1,0) & =A(\beta_{24})\\
\alpha_{62}= &(1,2,3,4,3,2,1,0) & =A(\beta_{21})\\
\alpha_{63}= &(2,2,3,4,3,2,1,0) & =A(\beta_{18})
\end{array}
$$
$$\begin{array}{cc@{\hspace{7mm}}c}
\beta_{0}= &   (0,0,0,0,0,0,0,1)  &   \gamma_{0}=(2,3,4,6,5,4,3,1)\\
\beta_{1}= &   (0,0,0,0,0,0,1,1)  &   \gamma_{1}=(2,3,4,6,5,4,2,1)\\
\beta_{2}= &   (0,0,0,0,0,1,1,1)  &   \gamma_{2}=(2,3,4,6,5,3,2,1)\\
\beta_{3}= &   (0,0,0,0,1,1,1,1)  &   \gamma_{3}=(2,3,4,6,4,3,2,1)\\
\beta_{4}= &   (0,0,0,1,1,1,1,1)  &   \gamma_{4}=(2,3,4,5,4,3,2,1)\\
\beta_{5}= &   (0,1,0,1,1,1,1,1)  &   \gamma_{5}=(2,2,4,5,4,3,2,1)\\
\beta_{6}= &   (0,0,1,1,1,1,1,1)  &   \gamma_{6}=(2,3,3,5,4,3,2,1)\\
\beta_{7}= &   (1,0,1,1,1,1,1,1)  &   \gamma_{7}=(1,3,3,5,4,3,2,1)\\
\beta_{8}= &   (0,1,1,1,1,1,1,1)  &   \gamma_{8}=(2,2,3,5,4,3,2,1)\\
\beta_{9}= &   (1,1,1,1,1,1,1,1)  &   \gamma_{9}=(1,2,3,5,4,3,2,1)\\
\beta_{10}= &   (0,1,1,2,1,1,1,1)  &   \gamma_{10}=(2,2,3,4,4,3,2,1)\\
\beta_{11}= &   (1,1,1,2,1,1,1,1)  &   \gamma_{11}=(1,2,3,4,4,3,2,1)\\
\beta_{12}= &   (0,1,1,2,2,1,1,1)  &   \gamma_{12}=(2,2,3,4,3,3,2,1)\\
\beta_{13}= &   (1,1,2,2,1,1,1,1)  &   \gamma_{13}=(1,2,2,4,4,3,2,1)\\
\beta_{14}= &   (1,1,1,2,2,1,1,1)  &   \gamma_{14}=(1,2,3,4,3,3,2,1)\\
\end{array}$$
$$\begin{array}{cc@{\hspace{7mm}}c}
\beta_{15}= &   (0,1,1,2,2,2,1,1)  &   \gamma_{15}=(2,2,3,4,3,2,2,1)\\
\beta_{16}= &   (1,1,2,2,2,1,1,1)  &   \gamma_{16}=(1,2,2,4,3,3,2,1)\\
\beta_{17}= &   (1,1,1,2,2,2,1,1)  &   \gamma_{17}=(1,2,3,4,3,2,2,1)\\
\beta_{18}= &   (2,2,3,4,3,2,1,1)  &   \gamma_{18}=(0,1,1,2,2,2,2,1)\\     
\beta_{19}= &   (1,1,2,3,2,1,1,1)  &   \gamma_{19}=(1,2,2,3,3,3,2,1)\\
\beta_{20}= &   (1,1,2,2,2,2,1,1)  &   \gamma_{20}=(1,2,2,4,3,2,2,1)\\
\beta_{21}= &   (1,2,3,4,3,2,1,1)  &   \gamma_{21}=(1,1,1,2,2,2,2,1)\\   
\beta_{22}= &   (1,2,2,3,2,1,1,1)  &   \gamma_{22}=(1,1,2,3,3,3,2,1)\\
\beta_{23}= &   (1,1,2,3,2,2,1,1)  &   \gamma_{23}=(1,2,2,3,3,2,2,1)\\
\beta_{24}= &   (1,2,2,4,3,2,1,1)  &   \gamma_{24}=(1,1,2,2,2,2,2,1)\\     
\beta_{25}= &   (1,2,2,3,2,2,1,1)  &   \gamma_{25}=(1,1,2,3,3,2,2,1)\\
\beta_{26}= &   (1,1,2,3,3,2,1,1)  &   \gamma_{26}=(1,2,2,3,2,2,2,1)\\
\beta_{27}= &   (1,2,2,3,3,2,1,1)  &   \gamma_{27}=(1,1,2,3,2,2,2,1)
\end{array}$$
$$\begin{array}{ccc}
\omega = &   (2,3,4,6,5,4,3,2)  & =A(\gamma_0)
\end{array}$$
\noindent Cubic form:
\begin{eqnarray*}
I_3&=&x_{1}x_{15}x_{18}+x_{1}x_{17}x_{21}+x_{1}x_{20}x_{24}-x_{1}x_{23}x_{27}+x_{1}x_{25}x_{26}+x_{2}x_{12}x_{18}+x_{2}x_{14}x_{21}+x_{2}x_{16}x_{24}\\&&
-x_{2}x_{19}x_{27}+x_{2}x_{22}x_{26}+x_{3}x_{10}x_{18}+x_{3}x_{11}x_{21}+x_{3}x_{13}x_{24}-x_{3}x_{19}x_{25}+x_{3}x_{22}x_{23}+x_{4}x_{8}x_{18}\\&&
+x_{4}x_{9}x_{21}+x_{4}x_{13}x_{27}-x_{4}x_{16}x_{25}+x_{4}x_{20}x_{22}-x_{5}x_{6}x_{18}-x_{5}x_{7}x_{21}+x_{5}x_{13}x_{26}-x_{5}x_{16}x_{23}\\&&
+x_{5}x_{19}x_{20}+x_{6}x_{9}x_{24}-x_{6}x_{11}x_{27}+x_{6}x_{14}x_{25}
-x_{6}x_{17}x_{22}-x_{7}x_{8}x_{24}+x_{7}x_{10}x_{27}-x_{7}x_{12}x_{25}\\&&
+x_{7}x_{15}x_{22}-x_{8}x_{11}x_{26}+x_{8}x_{14}x_{23}-x_{8}x_{17}x_{19}
+x_{9}x_{10}x_{26}-x_{9}x_{12}x_{23}+x_{9}x_{15}x_{19}-x_{10}x_{14}x_{20}\\&&
+x_{10}x_{16}x_{17}+x_{11}x_{12}x_{20}-x_{11}x_{15}x_{16}-x_{12}x_{13}x_{17}
\end{eqnarray*}
\noindent Cartan generators:
\begin{eqnarray*}
H_{\beta_0}&=&-y\pa+x_{0}\pa_{0}\\
H_{\alpha_1}&=&-x_{6}\pa_{6}+x_{7}\pa_{7}-x_{8}\pa_{8}+x_{9}\pa_{9}-x_{10}\pa_{10}+x_{11}\pa_{11}-x_{12}\pa_{12}+x_{14}\pa_{14}-x_{15}\pa_{15}+x_{17}\pa_{17}\\
&&+x_{18}\pa_{18}-x_{21}\pa_{21}\\
H_{\alpha_2}&=&-x_{4}\pa_{4}+x_{5}\pa_{5}-x_{6}\pa_{6}-x_{7}\pa_{7}+x_{8}\pa_{8}+x_{9}\pa_{9}-x_{19}\pa_{19}+x_{22}\pa_{22}-x_{23}\pa_{23}+x_{25}\pa_{25}\\
&&-x_{26}\pa_{26}+x_{27}\pa_{27}
\end{eqnarray*}
\begin{eqnarray*}
H_{\alpha_3}&=&-x_{4}\pa_{4}-x_{5}\pa_{5}+x_{6}\pa_{6}+x_{8}\pa_{8}-x_{11}\pa_{11}+x_{13}\pa_{13}-x_{14}\pa_{14}+x_{16}\pa_{16}-x_{17}\pa_{17}+x_{20}\pa_{20}\\
&&+x_{21}\pa_{21}-x_{24}\pa_{24}\\
H_{\alpha_4}&=&-x_{3}\pa_{3}+x_{4}\pa_{4}-x_{8}\pa_{8}-x_{9}\pa_{9}+x_{10}\pa_{10}+x_{11}\pa_{11}-x_{16}\pa_{16}+x_{19}\pa_{19}-x_{20}\pa_{20}+x_{23}\pa_{23}\\
&&+x_{24}\pa_{24}-x_{27}\pa_{27}\\
H_{\alpha_5}&=&-x_{2}\pa_{2}+x_{3}\pa_{3}-x_{10}\pa_{10}-x_{11}\pa_{11}+x_{12}\pa_{12}-x_{13}\pa_{13}+x_{14}\pa_{14}+x_{16}\pa_{16}-x_{23}\pa_{23}-x_{25}\pa_{25}\\
&&+x_{26}\pa_{26}+x_{27}\pa_{27}\\
H_{\alpha_6}&=&-x_{1}\pa_{1}+x_{2}\pa_{2}-x_{12}\pa_{12}-x_{14}\pa_{14}+x_{15}\pa_{15}-x_{16}\pa_{16}+x_{17}\pa_{17}-x_{19}\pa_{19}+x_{20}\pa_{20}-x_{22}\pa_{22}\\
&&+x_{23}\pa_{23}+x_{25}\pa_{25}\\
H_{\alpha_7}&=&-5-x_{0}\pa_{0}+x_{1}\pa_{1}-x_{15}\pa_{15}-x_{17}\pa_{17}-x_{18}\pa_{18}-x_{20}\pa_{20}-x_{21}\pa_{21}-x_{23}\pa_{23}-x_{24}\pa_{24}-x_{25}\pa_{25}\\
&&-x_{26}\pa_{26}-x_{27}\pa_{27}
\end{eqnarray*}
\noindent Simple roots:
\begin{eqnarray*}
E_{\alpha_{1}}&=&-x_{6}\pa_{7}-x_{8}\pa_{9}-x_{10}\pa_{11}-x_{12}\pa_{14}-x_{15}\pa_{17}+x_{21}\pa_{18}\\
E_{\alpha_{2}}&=&-x_{4}\pa_{5}-x_{6}\pa_{8}-x_{7}\pa_{9}+x_{19}\pa_{22}+x_{23}\pa_{25}+x_{26}\pa_{27}\\
E_{\alpha_{3}}&=&-x_{4}\pa_{6}-x_{5}\pa_{8}-x_{11}\pa_{13}-x_{14}\pa_{16}-x_{17}\pa_{20}+x_{24}\pa_{21}\\
E_{\alpha_{4}}&=&-x_{3}\pa_{4}+x_{8}\pa_{10}+x_{9}\pa_{11}+x_{16}\pa_{19}+x_{20}\pa_{23}+x_{27}\pa_{24}\\
E_{\alpha_{5}}&=&-x_{2}\pa_{3}+x_{10}\pa_{12}+x_{11}\pa_{14}+x_{13}\pa_{16}+x_{23}\pa_{26}+x_{25}\pa_{27}\\
E_{\alpha_{6}}&=&-x_{1}\pa_{2}+x_{12}\pa_{15}+x_{14}\pa_{17}+x_{16}\pa_{20}+x_{19}\pa_{23}+x_{22}\pa_{25}\\
E_{\alpha_{7}}&=&-x_{0}\pa_{1}+i(-x_{15}x_{18}-x_{17}x_{21}-x_{20}x_{24}+x_{23}x_{27}-x_{25}x_{26})/y\\
E_{-\alpha_{1}}&=&x_{7}\pa_{6}+x_{9}\pa_{8}+x_{11}\pa_{10}+x_{14}\pa_{12}+x_{17}\pa_{15}-x_{18}\pa_{21}\\
E_{-\alpha_{2}}&=&x_{5}\pa_{4}+x_{8}\pa_{6}+x_{9}\pa_{7}-x_{22}\pa_{19}-x_{25}\pa_{23}-x_{27}\pa_{26}\\
E_{-\alpha_{3}}&=&x_{6}\pa_{4}+x_{8}\pa_{5}+x_{13}\pa_{11}+x_{16}\pa_{14}+x_{20}\pa_{17}-x_{21}\pa_{24}\\
E_{-\alpha_{4}}&=&x_{4}\pa_{3}-x_{10}\pa_{8}-x_{11}\pa_{9}-x_{19}\pa_{16}-x_{23}\pa_{20}-x_{24}\pa_{27}\\
E_{-\alpha_{5}}&=&x_{3}\pa_{2}-x_{12}\pa_{10}-x_{14}\pa_{11}-x_{16}\pa_{13}-x_{26}\pa_{23}-x_{27}\pa_{25}\\
E_{-\alpha_{6}}&=&x_{2}\pa_{1}-x_{15}\pa_{12}-x_{17}\pa_{14}-x_{20}\pa_{16}-x_{23}\pa_{19}-x_{25}\pa_{22}\\
E_{-\alpha_{7}}&=&x_{1}\pa_{0}+iy(\pa_{15}\pa_{18}+\pa_{17}\pa_{21}+\pa_{20}\pa_{24}-\pa_{23}\pa_{27}+\pa_{25}\pa_{26})
\end{eqnarray*}


\newpage

\providecommand{\href}[2]{#2}\begingroup\raggedright\endgroup

\end{document}